\documentclass[useAMS]{mn2e}

\usepackage{graphicx}

\title[Future supernovae data and quintessence models]
{Future supernovae data and quintessence models}
\author[Elisa Di Pietro and Jean-Fran{\c c}ois Claeskens]{Elisa 
Di Pietro$^{1}$\thanks{E-mail\,: dipietro@astro.ulg.ac.be} and 
Jean-Fran{\c c}ois Claeskens$^1$\thanks{E-mail\,: 
claeskens@astro.ulg.ac.be} \\
$^{1}$Institut d'Astrophysique et de G\'eophysique,
Universit\'e de Li\`ege, All\'ee du 6 Ao\^ut 17, B-4000 LIEGE-BELGIUM}
\begin{document}

\date{December 2002}

\pagerange{\pageref{firstpage}--\pageref{lastpage}} \pubyear{2002}

\maketitle

\label{firstpage}

\begin{abstract}
The possibility to unambiguously determine the equation-of-state
of the cosmic dark energy with existing and future supernovae data 
is investigated. We consider four evolution laws for this 
equation-of-state corresponding to four quintessential models, i.e. 
i) a cosmological constant, ii) a general barotropic fluid, iii) a 
perfect fluid with a linear equation-of-state and iv) a more physical 
model based on a pseudo-Nambu-Goldstone boson field. We explicitly 
show the degeneracies present not only within 
each model but also between the different models\,: they are caused 
by the multi-integral relation between the equation-of-state of dark 
energy and the luminosity distance. 
Present supernova observations are analysed using a standard $\chi^2$ 
method and the minimal $\chi^2$ values obtained for each model are compared. 
We confirm the difficulty to discriminate between these models using present
SNeIa data only.
By means of simulations, we then show that future SNAP observations will not
remove all the degeneracies. For example, wrong estimations of $\Omega_m$ with 
a good value of $\chi^2_{min}$ could be found if the right cosmological model
is not used to fit the data.
We finally give some probabilities to obtain unambiguous
results, free from degeneracies. In particular, the probability 
to confuse a cosmological constant with a true barotropic fluid with an 
equation-of-state different from $-1$ is shown to be $95\,\%$ at a $2\,\sigma$ level.
\end{abstract}


\maketitle  

\section{Introduction}

Present supernovae data strongly support cosmological models containing
a perfect fluid with a negative pressure (Riess et~al. 1998, Perlmutter et
al. 1999). The oldest and most studied candidate for this fluid is
the cosmological constant, which acts like a perfect fluid whose
equation-of-state (hereafter, EOS) is $w \equiv p\,/\,\rho = -1$, where 
$p$ is the fluid pressure and $\rho$ its density. But the vacuum energy 
density associated with the cosmological constant is 60-120 orders of 
magnitude smaller than its natural value derived from quantum field theories.
This discrepancy is known as the cosmological constant problem (Abbott
1988, Weinberg 1989, Carroll et~al. 1992, Sahni \& Starobinsky 2000) and 
has led theorists to find alternative dark energy candidates with a present 
negative pressure. The simplest models are based on a generalisation of the 
cosmological constant, e.g. a barotropic fluid ($w(z) = w_0 =$ constant $< 
1$\,; Gonz\'alez-D\'\i az 2000, Di Pietro \& Demaret 2001) or a homogeneous 
fluid for which $w$ is a linear function of $z$ ($w(z) = w_0 + w_1\,z$ 
with $w_0$ and $w_1$ constant\,; Goliath et~al. 2001, Maor et al. 2001, 2002).

These models are interesting because they are described by simple field
equations. However, they suffer from a lack of physical justification.
Other models with a more general EOS and stronger physical
interpretation have been proposed (Peebles \& Ratra 1988, Ratra \&
Peebles 1988, Wetterich 1988, Ferreira \& Joyce 1998, Steinhardt et~al.
1999). The most popular ones are based on a dynamical quintessential component 
represented by a minimally coupled scalar field evolving in a potential. 
Mathematically, such a quintessence fluid can be described by an EOS $w$ 
function of the redshift $z$ ($-1 \leq w(z) \leq 1$). Among these 
quintessence models, we shall consider the one which assumes the existence
of an ultra-light pseudo-Nambu Goldstone boson (PNGB) field relaxing to
its vacuum state (Frieman \& Waga 1998, Waga \& Frieman 2000, Ng \&
Wiltshire 2001a). From the quantum viewpoint, the PNGB models are the
simplest way to introduce a naturally ultra-light scalar field 
able to reproduce the cosmological observations (Frieman et~al. 1992,
1995). Moreover PNGB models provide an interesting theoretical framework
to any spontaneous symmetry breaking which could justify the neutrino mass
found in the Mikheyev-Smirnov-Wolfenstein solution to the solar neutrino
problem (Wolfenstein 1979, Mikheyev \& Smirnov 1985).

In summary, many theoretical models describing a cosmological fluid with
negative pressure have been proposed\,: some are designed to be
mathematically simple while others rely on a more physical justification. 
The first aim of this paper is to put in evidence the strong degeneracies
existing between the luminosity distance predicted by four quintessence 
models\,: a cosmological constant, a
general barotropic fluid, a fluid with a linear equation-of-state
and a more physical model based on a pseudo-Nambu-Goldstone boson field.
Second, we shall determine the constraints on those models 
coming from the present observations of type Ia supernovae (hereafter, SNeIa)
and confirm the need for more data to discriminate between
them. This may be fulfilled by the proposed SNAP satellite
(SuperNova/Acceleration Probe\,; see SNAP URL). The
main objective of this instrument is to detect a very large number of
supernovae up to a redshift of $1.7$, in order to yield a more
precise determination of the cosmological parameters and therefore to
provide information on the nature of dark energy. The third objective 
of this paper is to analyse, by means of simulated data, how well 
future SNAP observations alone will be able to break the degeneracies 
that we put in evidence. 

Several authors have already studied the feasibility 
of SNAP to determine the properties of the dark energy. Depending on the
method used for reconstructing the cosmological model, their conclusions are
quite different\,: some authors are optimistic regarding the possible 
determination of the EOS of the dark energy using SNeIa (Huterer \& Turner
1999, Nakaruma \& Chiba 1999, Saini et~al 2000, Chiba \& Nakaruma 2000, 
Weller \& Albrecht 2001, 2002) while others are more cautious (Barger \&
Marfatia 2001, Astier 2001, Maor et~al. 2001, 2002, Gerke \& Efstathiou
2002). Most of these discrepancies can be traced back to the differences in
initial assumptions, prior knowledge, ... used for the
reconstruction of the cosmological model (Goliath 2001). Usually, optimistic
conclusions result from strong assumptions, such as an accurate knowledge
of $\Omega_m$. 

The structure of the paper is as follows. In Section~2, we present the field 
equations describing the four cosmological models we have considered. In Section 3, 
we explicitly display the degeneracies between the luminosity distances predicted 
by these four models. The constraints brought by present SNeIa data on the parameters
of each model are 
presented in Section~4. In Section~5, we explore the ability of SNAP to break 
the degeneracies and so to discriminate among these models. 
Finally, we summarize our work and give some conclusions in Section~6.

\section{Quintessence models\,: theoretical framework}

The field equations of a Friedmann-Lema\^\i tre-Robertson-Walker 
spacetime filled with two non-interacting fluids, a pressureless fluid
(matter - $\Omega_m$) and a spatially homogeneous scalar field $\phi$
in a potential $V(\phi)$ (quin\-tes\-sen\-ce - $\Omega_x$), can be written
as
\begin{eqnarray}
\displaystyle 
\frac{H^2(z)}{H_0^2} & = & \Omega_m \,(1 + z)^3 + \Omega_k\,(1 + z)^2
+ \Omega_x\,g(z)  \label{hubble}\\
\displaystyle
\rho'_x(z) & = & -\,3\,[\rho_x(z) + p_x(z)]\,H(z)  \nonumber \\
& = & \,-\,3\,[1 + w(z)]\,\rho_x(z)\,H(z),
\label{phi}
\end{eqnarray}
where the prime denotes the time derivative, $H(z)$, the Hubble parameter,
$\Omega_k$, the curvature parameter ($\Omega_k = 1 - \Omega_m - \Omega_x$)
and where
\begin{eqnarray}
\displaystyle \kappa\,\rho_x(z) & \equiv & \displaystyle
\frac{1}{2}\,\phi'^2(z) + \,V(\phi(z)), \label{rhophi} \\
\displaystyle \kappa\,p_x(z) & \equiv & \displaystyle
\frac{1}{2}\,\phi'^2(z) - \,V(\phi(z)), \\
w(z) & \equiv & \displaystyle \frac{\,p_x(z)}{\,\rho_x(z)}, \\
g(z) & \equiv & \displaystyle \frac{\rho_x (z)}{\rho_{x,0}} \\
& = & \displaystyle
(1 + z)^3\,\,\exp\left[\int_0^z{3\,w(z')\,d\,\ln(1+z')}\right],
\label{functiong}
\end{eqnarray}
with $\kappa = 8\,\pi\,G$.

When the scalar field acts like a fluid with a linear EOS in $z$, i.e.
$w(z) = w_0 + w_1\,z$, equation (\ref {functiong}) becomes
$g(z) = e^{3\,w_1\,z}\,\, (1 + z)^{3\,(1 + w_0 - w_1)}$ and the Friedmann
equation given by (\ref {hubble}) can be written as
\begin{eqnarray}
H^2(z) & = & \displaystyle
H_0^2\,\left\{(1 + z)^2 \left[ \Omega_m\,z\, + 1 \right] \right.\nonumber \\
& + & \displaystyle \left. \Omega_x (1 + z) ^2 
\left[ e^{3 w_1 z}\,(1 + z)^{1 + 3 (w_0 - w_1)} - 1 \right]
\right\},
\end{eqnarray}
where the relation between the density parameters has been used, i.e.
$\Omega_k = 1 - \Omega_m - \Omega_x$.

The luminosity distance $d_L(z)$ to an object at redshift $z$ is such that 
\begin{eqnarray}
\displaystyle
D_L^{th}(z) & \equiv & \displaystyle \frac{d_L(z)\,H_0}{c} \\
& \equiv & \displaystyle
\frac{(1+z)}{\sqrt{\mid \Omega_k\mid }} \,\,\,\,
S\left( \int_0^z \frac{\sqrt{\mid \Omega_k \mid}\,\,H_0\,\, dz'}{H(z')} 
\right) 
\label{dl}
\end{eqnarray}
with $S(x) = \sin (x), x, \sinh (x)$ for respectively a spatially 
closed, flat and open universe. Note that $d_L$ is the luminosity 
distance whereas $D^{th}_L$ is a dimensionless quantity. For $d_L$ in units 
of Mpc, the apparent magnitude can be expressed in terms of $D_L^{th}$ as 
follows 
\begin{equation}
\displaystyle
m_B = 5\,\log(D_L^{th}) + \bar{M}_B
\end{equation}
where $\bar{M}_B$ is the magnitude zero point in the $B$ band defined by
\begin{equation}                                                       
\displaystyle
\bar{M}_B \equiv M_B - 5\, \log [ H_0\,/\,c\, ] + 25
\label{barm}
\end{equation}
with $M_B$, the absolute magnitude of the SNeIa.

Hereafter, we focus on three types of models
containing a scalar fluid with a linear EOS (cf. also models A, B and C 
in Table~\ref {tablemodel})\,:
\begin{enumerate}
\item[- ] \underline{Model A}\,: a scalar field still acting as a
cosmological constant with no constraint on the geometry of the universe, 
i. e. a FLRW model,
\item[- ] \underline{Model B}\,: a scalar fluid with a {\em constant} EOS in a 
spatially flat universe,
\item[- ] \underline{Model C}\,: a scalar fluid with a {\em linear} EOS in a
spatially flat universe.
\end{enumerate}

\begin{figure}
\begin{center}
\includegraphics[height=16pc,width=19pc]{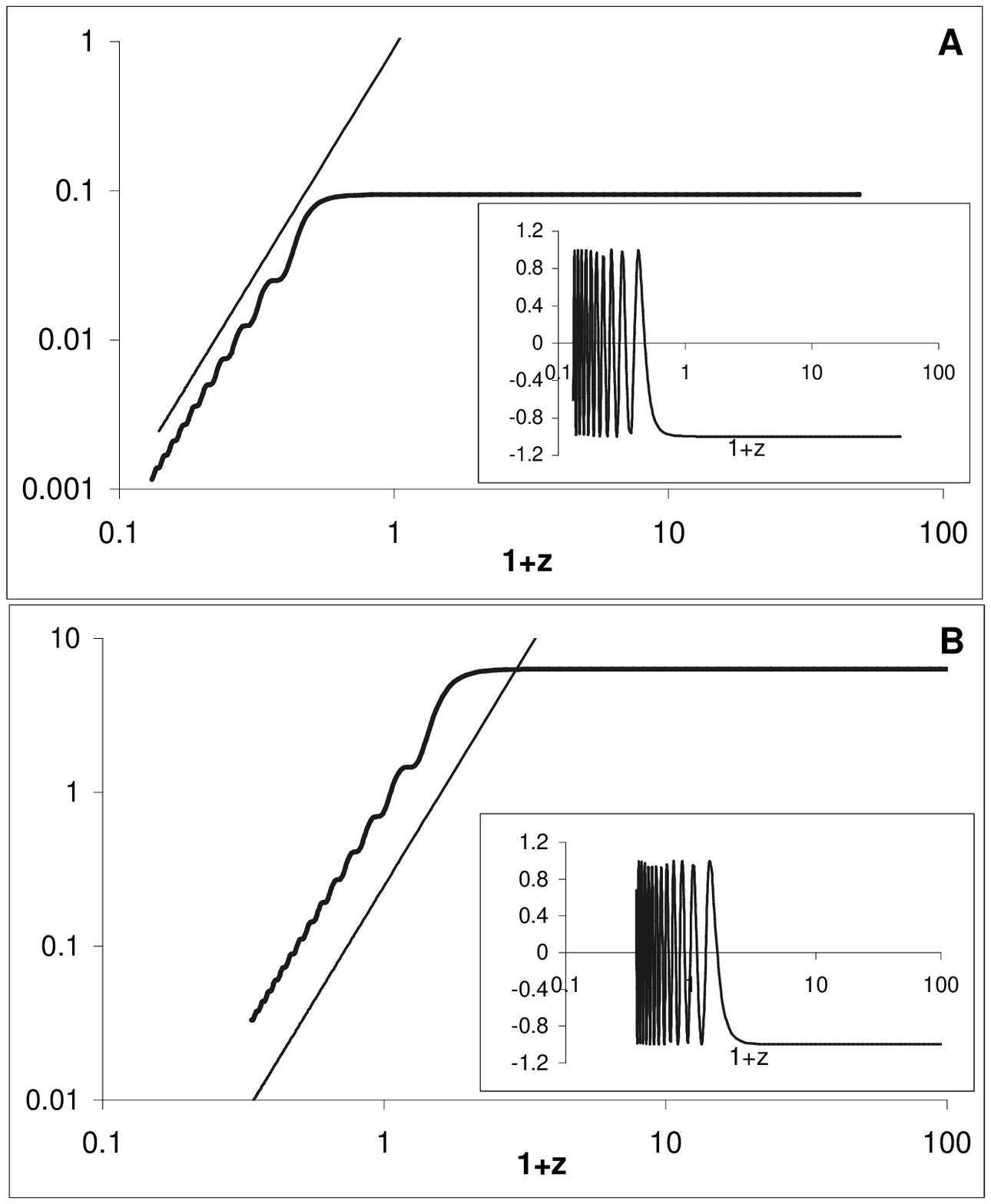}
\caption{Behavior of $\rho_\phi$ (thick) and $\rho_m$ (thin), in units 
of the present critical density, when 
the quintessence component is a PNGB field. In the small boxes, the 
corresponding behavior of $w_\phi$ is given. Negative redshifts are 
associated with the future of the universe. \underline{For A\,:} $\,M = 
1.4 \times 10^{-3} h_0^{1/2}$ eV and $\,f = 3 \times 10^{17}$ GeV.
\underline{For B\,:} $\,M = 4 \times 10^{-3} h_0^{1/2}$ eV and $\,f 
= 5 \times 10^{17}$ GeV.}
\label{fig1}
\end{center}
\end{figure}

\begin{figure}
\begin{center}
\includegraphics[height=7.5cm]{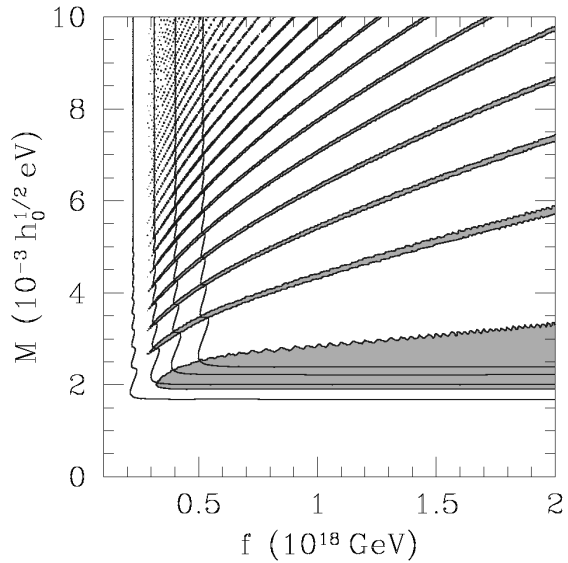}
\caption{
$M-f$ diagram for the PNGB model. The shaded areas represent the parameter
regions where the universe is presently accelerating whereas the solid
lines are the curves of equal $\Omega_m$ with $\Omega_m = 0.8$, $0.6$,
$0.4$, $0.2$, from left to right.}
\label{figclassifd}
\end{center}
\end{figure}

These models are mathematically the simplest ones but they have no 
satisfactory physical interpretation. So we shall also consider a fourth 
quintessence model, based on physical arguments. 
This model is constructed with an ultra-light pseudo-Nambu-Goldstone 
boson (PNGB) field (Frieman \& Waga 1998, Waga \& Frieman 2000, Ng \&
Wiltshire 2001a, 2001b). In quantum field theory, a Nambu-Goldstone boson
field is generated when a global symmetry is spontaneously broken.
Moreover, if this symmetry is explicitly broken, the Nambu-Goldstone
boson acquires a mass and becomes a pseudo-Nambu-Goldstone boson. A
quintessence model based on a PNGB field can be described by the field
equations (\ref {hubble})-(\ref {functiong}) with the following periodic
potential (Frieman et~al. 1992, 1995)\,:
\begin{equation}
\displaystyle
V(\phi) = M^4\,\left( \cos\left[\phi\,/\,f\right] + 1 \right)
\end{equation}
where $M$ and $f$ are the two mass scales, i.e. $f$ is the spontaneous-
and $M$ is the explicit symmetry breaking scale. For such a model, the 
field equations (\ref {hubble})-(\ref {functiong}) have no analytical 
solution. Therefore, the behavior of the functions
$g(z)$ and $H(z)$ can only be derived from a numerical integration. We take
as initial conditions $\phi'_i = 0$ and $\phi_i = 10^{-2}\, m_{Pl}$ where
$m_{Pl}$ is the reduced Planck mass (see justification below). Assuming
again a spatially flat universe, we are left with a model depending on only
two parameters, $M$ and $f$. This model is referred to as ``\underline{Model
D}\,'' (cf. also Table~\ref {tablemodel}) and its most important 
properties are summarized in Figures~\ref {fig1} and \ref {figclassifd}. 
Because of the Hubble damping, the PNGB field is
initially frozen near the top of its potential ($\,\phi_i' \sim 0\,$ and
$\,\phi_i \ll\,f$) and therefore acts as a cosmological constant, so that 
$w(z \gg 1) \approx - 1$. When the
universe temperature becomes less than the PNGB mass, i.e. $\,m_\phi =
M^2\,/\,f\,$, the field starts to slowly roll down the potential.
The minimum of the potential is reached asymptotically, after a large
number of coherent oscillations about it. This second phase corresponds
to a scalar component which behaves on average as nonrelativistic matter.
It means that model D looks initially like model A with $\Omega_k = 0$ 
and asymptotically like an Einstein-de Sitter universe. The larger the
quintessence mass, the sooner the asymptotic state
is reached. If the scalar field evolution starts before its energy density
becomes dominant, the PNGB field will never be relevant for the universe
dynamics (see Figure~\ref {fig1}A). However, if the dynamical 
phase of the scalar field begins when the vacuum energy associated with the PNGB 
field is already dominant, then the scalar energy density may dominate today the 
cosmic density of the universe and provide a non-negligible dark energy component
(see Figure~\ref {fig1}B).

\begin{table}
\begin{center}
\begin{tabular}{|c|l|c|}
\hline
Name & Constraints & Parameters \\
\hline
Model A & $w_0 = \,-\,1$ and $w_1 = 0$ &
$\Omega_m$, $\Omega_x$ \\
Model B & $\Omega_m + \Omega_x = 1$ and $w_1 = 0$
& $\Omega_m$, $w_0$ \\
Model C & $\Omega_m + \Omega_x = 1$ and $\Omega_m = 0.3$ & $w_0$, $w_1$ \\ 
Model D & $\Omega_m + \Omega_x = 1$ and & $M$, $f$ \\
& $V(\phi) = M^4\,(\cos[\phi\,/\,f] + 1)$ & \\
\hline
\end{tabular}
\caption{Characteristics of the four theoretical models considered
in this paper.}
\label{tablemodel}
\end{center}
\end{table}

Figure~\ref {figclassifd} shows examples of 
curves of equal $\Omega_m$ in an $M-f$ diagram.
The shaded area represents the parameter regions in which 
the expansion is accelerating today. In these regions, the kinetic energy of 
the PNGB field is small compared to its potential. This occurs either 
when the scalar field is still frozen (the lowest and largest shaded band) 
or when its first derivative changes sign, i.e. when $\phi' \sim 0$ (the 
thin shaded bands). Regions outside these shaded areas correspond to models where
the scalar field is presently near the minimum of its potential (for more
details on PNGB models, see Frieman et~al. 1992, 1995, Frieman \& Waga
1998, Waga \& Frieman 2000, Ng \& Wiltshire 2001a). 
\begin{figure*}
\begin{center}
\begin{tabular}{ll}
\includegraphics[width=13pc]{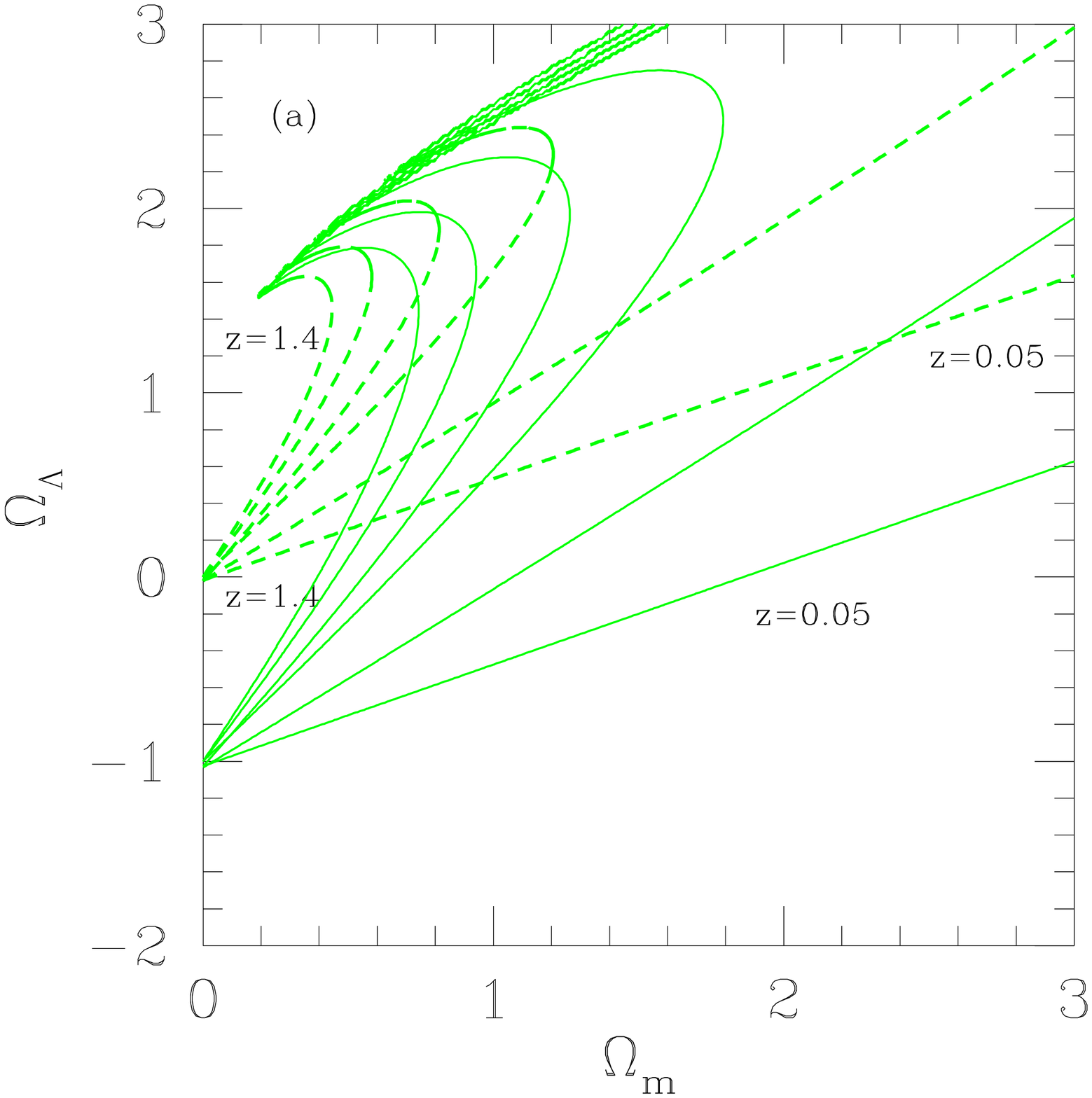} \hspace*{8mm}
\includegraphics[width=13pc]{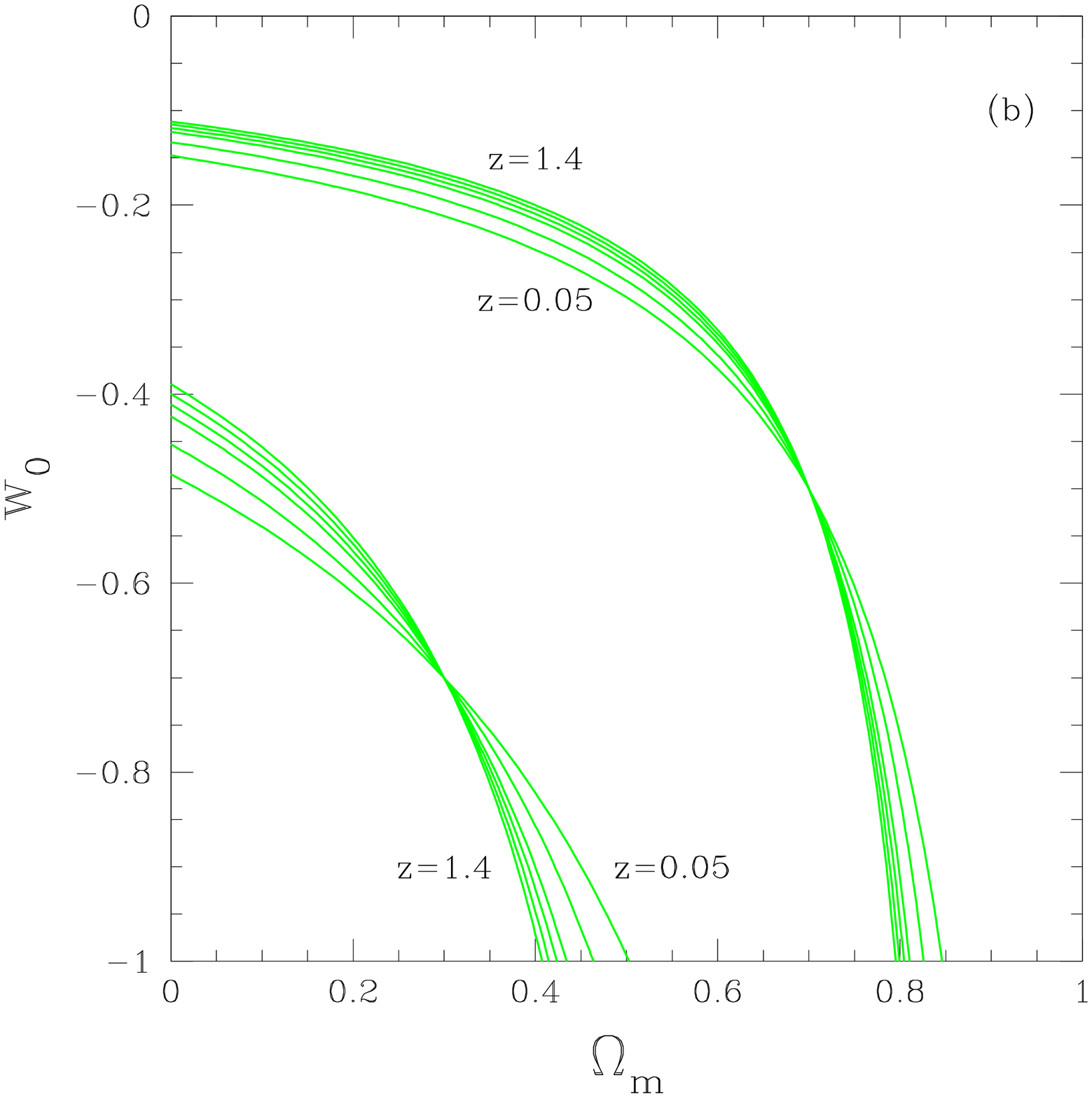} \\
\includegraphics[width=13pc]{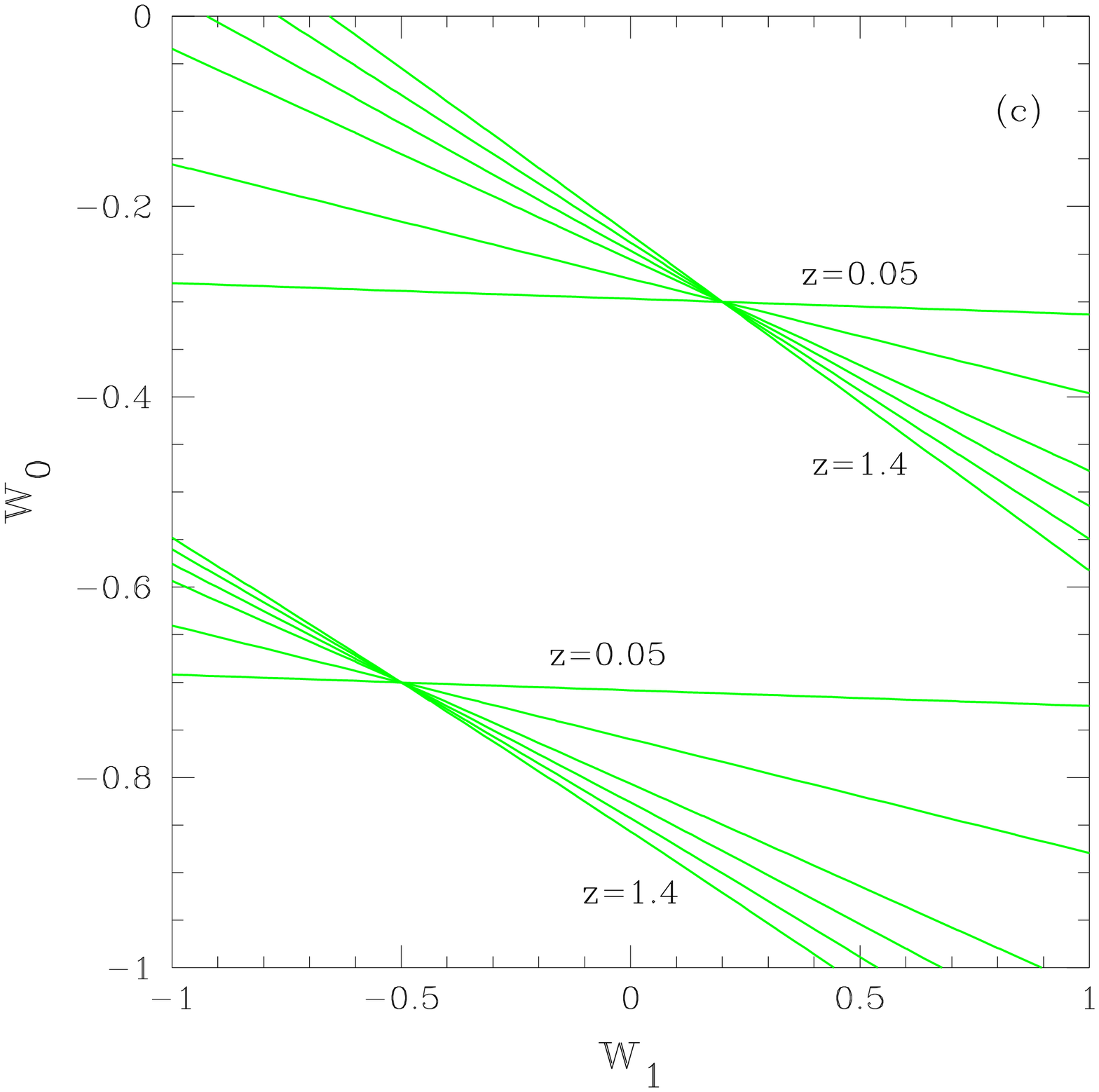} \hspace*{8mm} 
\includegraphics[width=13pc]{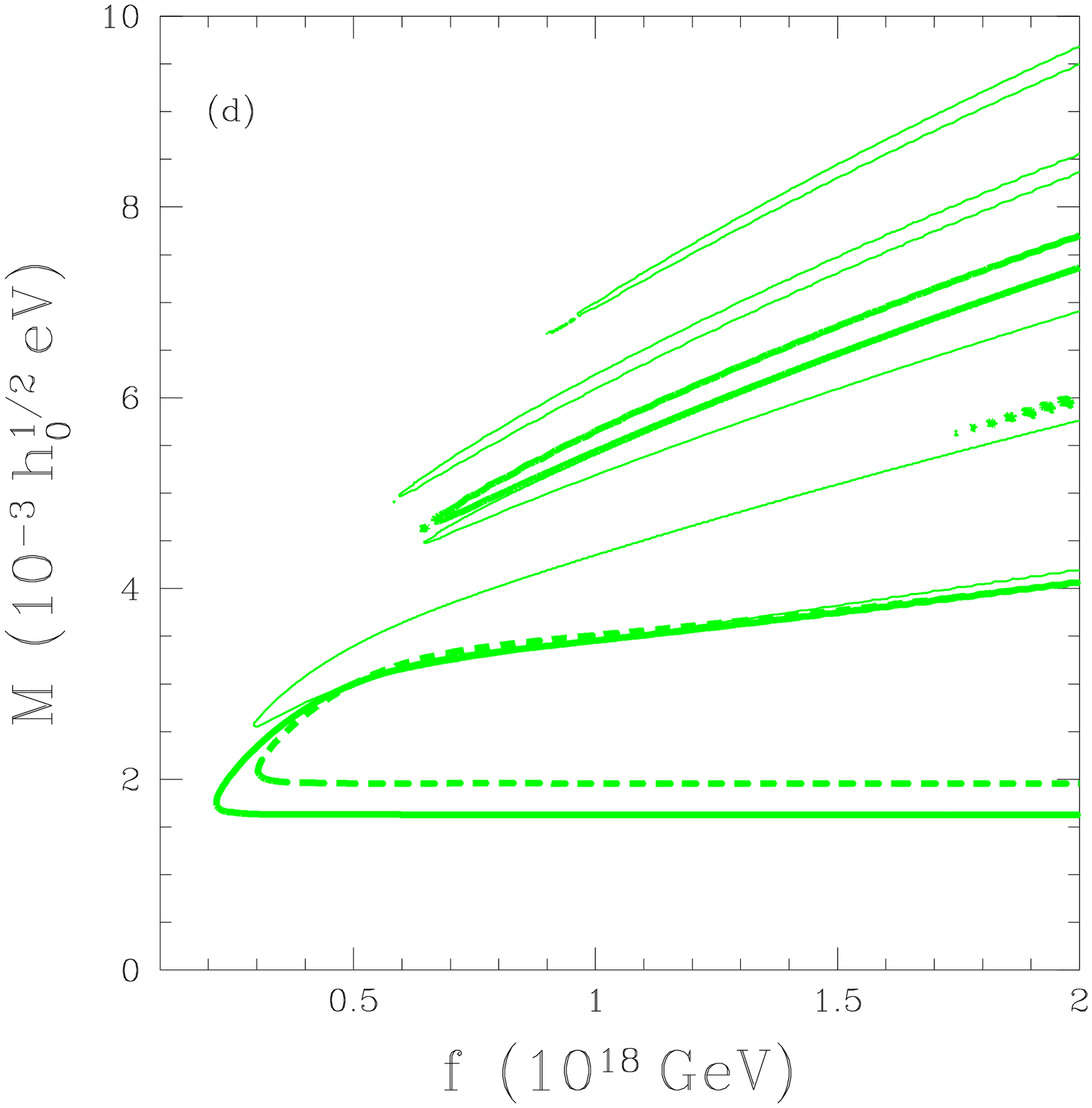}
\end{tabular}
 \caption{
Curves of equal luminosity distance in the framework of the four 
cosmological models considered in this paper. 
(a) \underline{Model A\,:} the dashed curves represent 
the set of values for $(\Omega_m, \Omega_\Lambda)$ such that 
$D_L(z_i; \Omega_m, \Omega_\Lambda)=D_L(z_i; 0, 0)$, with $z_i = 
0.05, 0.4, 0.8, 1.0, 1.2, 1.4$. Similarly, the solid curves show the 
set of values for $(\Omega_m, \Omega_\Lambda)$ such that 
$D_L(z_i; \Omega_m, \Omega_\Lambda)=D_L(z_i; 0, -1)$, with $z_i = 
0.05, 0.4, 0.8, 1.0, 1.2, 1.4$. (b) \underline{Model B\,:} 
the curves represent the set of values for $(\Omega_m, w_0)$ such 
that $D_L(z_i; \Omega_m, w_0) = D_L(z_i; 0.7, -0.5)$ (top) and 
$D_L(z_i; \Omega_m, w_0)=D_L(z_i; 0.3, -0.7)$ (bottom), with $z_i 
= 0.05, 0.4, 0.8, 1, 1.2, 1.4$. (c) \underline{Model C\,:} the 
curves represent the set of values for $(w_0, w_1)$ such that 
$D_L(z_i; w_0, w_1) = D_L(z_i; -0.3, 0.2)$ (top) and $D_L(z_i; w_0, 
w_1) = D_L(z_i; -0.7, -0.5)$ (bottom), with $z_i = 0.05, 0.4, 0.8, 
1, 1.2$ and $1.4$. (d) \underline{Model D\,:} curves show the set 
of values of $(f, M)$ such that $D_L(z_i; f, M) = D_L(z_i; 0.5, 3)$ 
with $z_i = 0.4$ (thin), $0.8$ (thick) and $1.2$ (dashed).}
\label{figdeg4}
\end{center}
\end{figure*}

\section{Degeneracies between the quintessence models}
\label{degeneracies}
Since the SNeIa apparent 
magnitudes are sensitive to the luminosity distance and since 
the latter is related to the model parameters through a multiple integral
(see eq. (\ref {dl}) with (\ref {hubble}) and (\ref {functiong})), it is quite 
natural to expect some degeneracies between the different models. This is even
true within a given cosmological model. Indeed, Figure~\ref {figdeg4} shows 
curves of equal luminosity distance for different redshifts, in the 
parameter spaces of the four models. For each model, the curve shape varies 
with the given redshift. The curves intersect in one point in models A, B and C and this 
lifts the degeneracy. But in model D, the contours could remain close to 
each other, making more difficult a clear estimation of the cosmological parameters.

\begin{figure}
\begin{center}
\begin{tabular}{c}
\includegraphics[height=10pc,width=18pc]{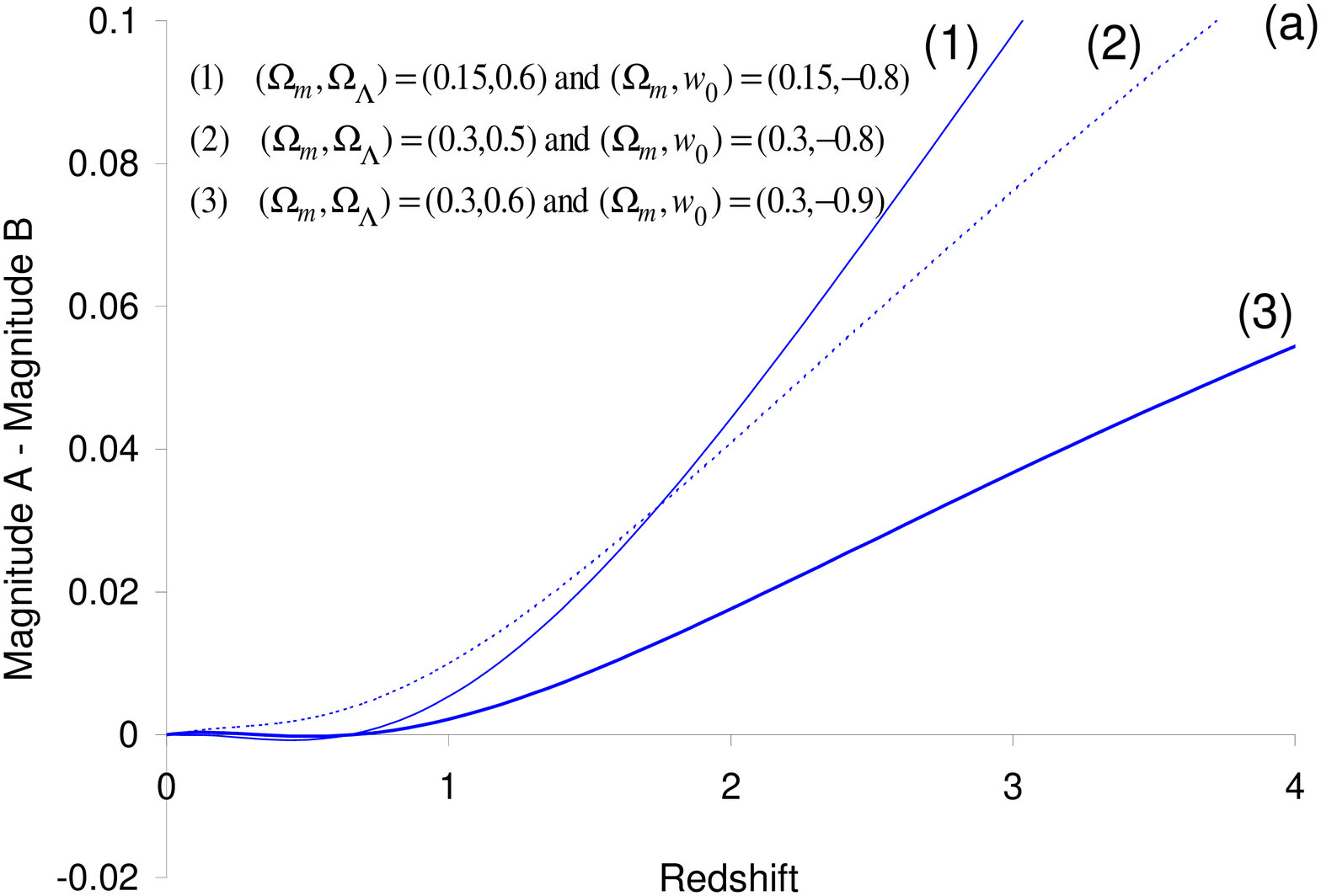} \\
\includegraphics[height=10pc,width=18pc]{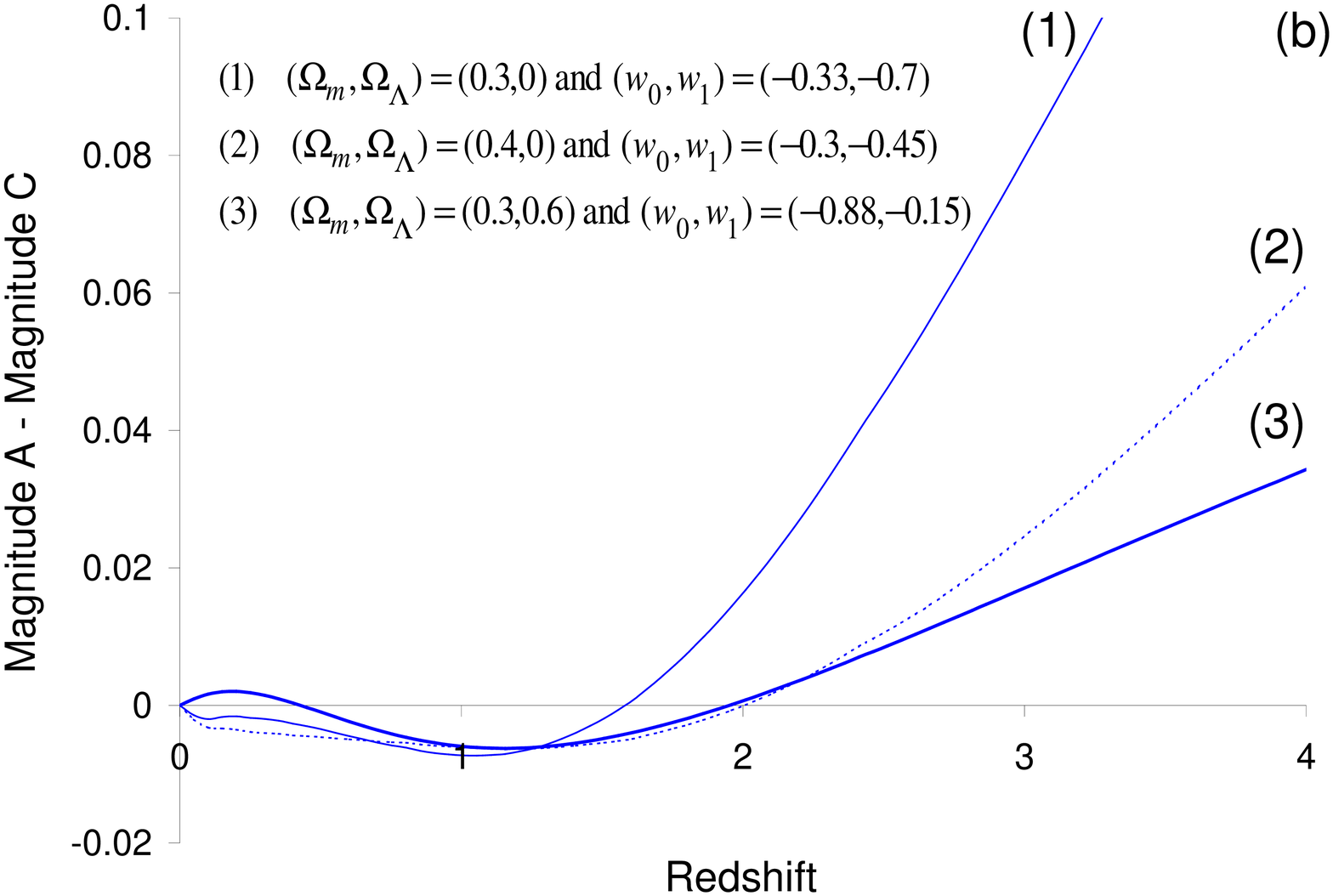} \\
\includegraphics[height=10pc,width=18pc]{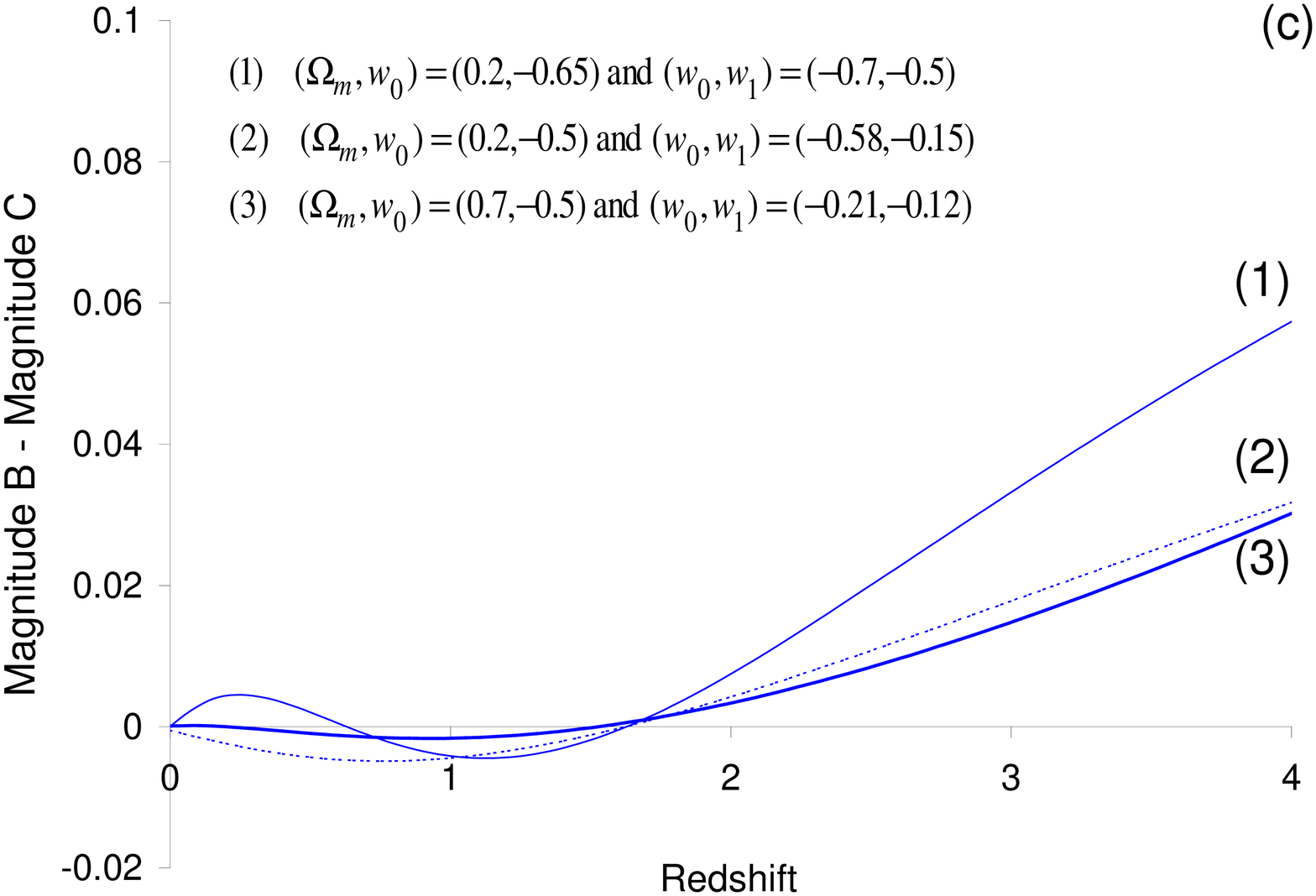} \\
\includegraphics[height=10pc,width=17.8pc]{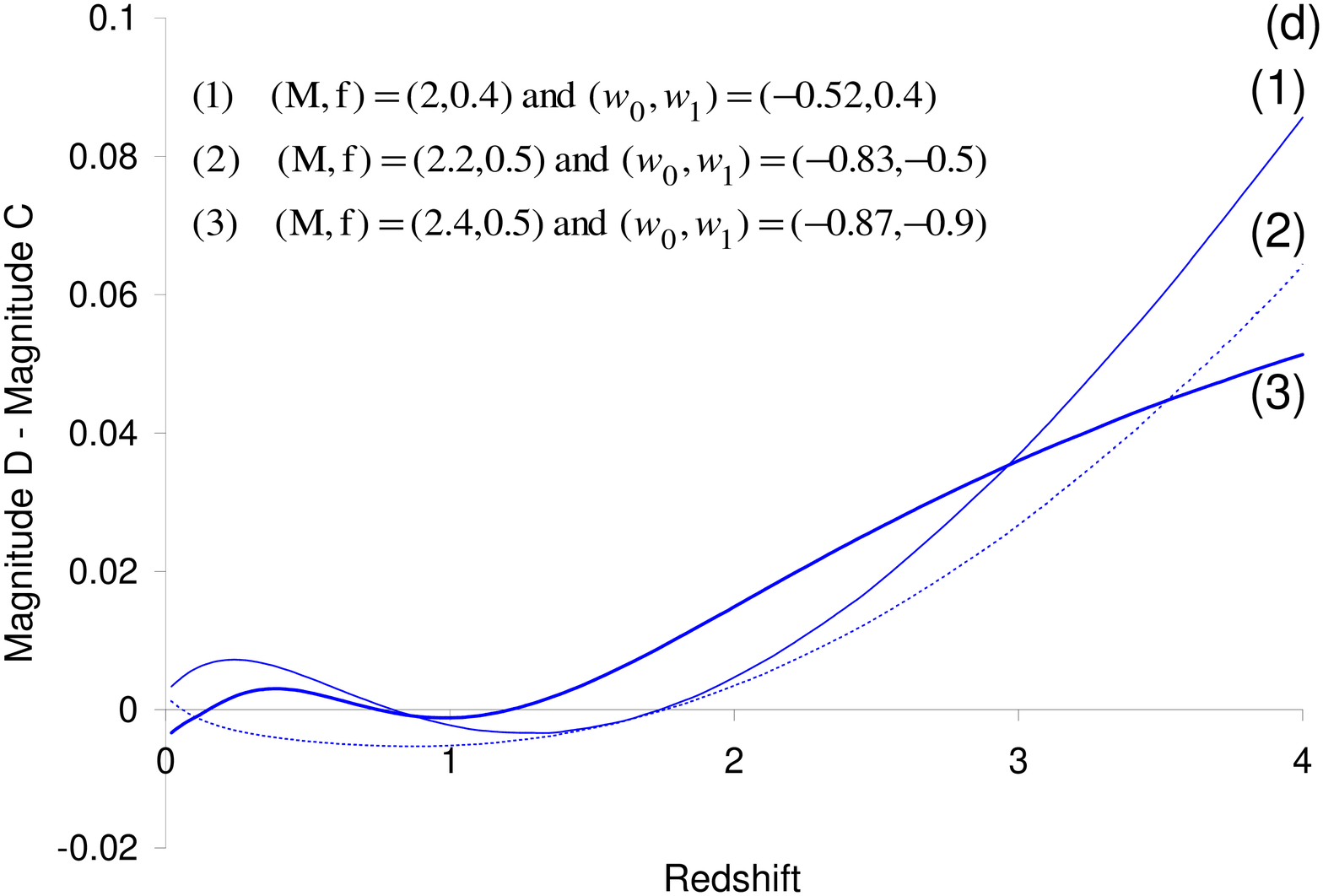}
\end{tabular}
\caption{Comparison between the magnitudes predicted by various models A
and B {\em (a)}, models A and C {\em (b)}, models B and C {\em (c)} and
models C and D {\em (d)}.}
\label{magnfig}
\end{center}
\end{figure}

The situation is even worse when comparing
the different models. Indeed, Figure~\ref {magnfig} shows expected 
magnitude differences between different models, versus redshift. Specific choices of 
parameters of either model A, B, C or D can yield magnitude 
differences less than 0.04 mag all the way to $z = 2$. These are really small 
compared to the 0.15 mag intrinsic dispersion of the SNeIa maximum 
luminosity. 
In other words, as already mentionned by Maor et al. (2001, 2002),
different function $w(z)$ can approximatively lead to the same $d_L(z)$ relation.
A large number of SNeIa observations with very good photometry are needed in 
order to break these degeneracies. At this point, data at large redshifts
($z > 2$) could be more than useful. 

\begin{figure}
\begin{center}
\includegraphics[width=15pc]{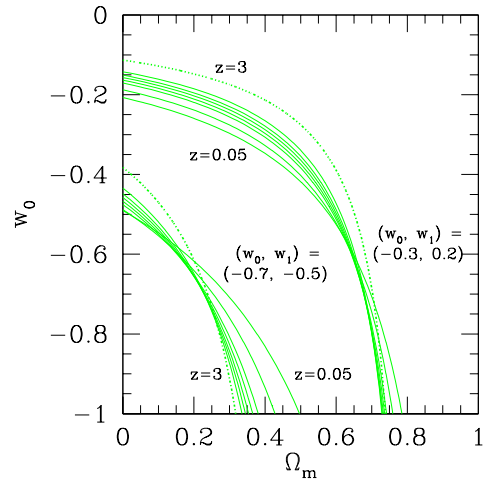} 
\caption{Curves in the parameter space of \underline{model B}
such that $D_L(z_i; \Omega_m, w_0)$ built with this model is equal 
to $D_L(z_i; w_0 = -0.7, w_1 = -0.5)$ (left) and to $D_L(z_i; w_0 
= -0.3, w_1 = 0.2)$ (right) derived from \underline{model C}, for 
$z_i = 0.05, 0.4, 0.8, 1, 1.2, 1.4, 1.7, 3$. The presence of 
degeneracies is revealed 
by the fact that the different curves cross each other over a very small 
area in the parameter space. The curves corresponding to $z = 3$ (dotted)
indicate that the degeneracy could break at very large redshifts.}
\label{figdeg1}
\end{center}
\end{figure}

Figure~\ref {figdeg1} illustrates that estimates of the cosmological parameters
can simply be erroneous if a wrong cosmological model is chosen for the analysis. 
Here are the different steps followed to draw this plot\,:
\begin{enumerate}
\item[$\star$] We chose some values of the parameters describing the model C, 
i.e. $(w_0, w_1) = (-0.7, -0.5)$.
\item[$\star$] The luminosity distance predicted by \underline{model C} 
at $z = 0.05$ were numerically computed.
\item[$\star$] We then searched the parameters $(\Omega_m, w_0)$
for which \underline{model B} predicts the same luminosity distance at 
$z = 0.05$. We plotted the curve of equal luminosity
distance obtained in the parameter space of model B.
\item[$\star$] We repeated the two previous steps for different values
of $z$, i.e. $z = 0.4, 0.8, 1, 1.2, 1.4, 1.7, 3$, and plotted the $6$
corresponding curves.
\item[$\star$] The four previous operations were repeated with a
second set of parameters for the model C, i.e. $(w_0, w_1) = (-0.3, 0.2)$.
\end{enumerate}

Keeping in mind that models C are defined with $\Omega_m = 0.3$, 
Figure 5 clearly shows that the value of 
$\Omega_m$ would not be properly recovered by fitting model B on the 
data. Indeed, the first set of curves ( correspondong to 
$(w_0, w_1) = (-0.7, -0.5)$ in model C) all cross over a very small area 
characterized by $(\Omega_m, w_0) \approx (0.15, 
-0.6)$. Similarly, the second set of curves (corresponding to 
$(w_0, w_1) = (-0.3, 0.2)$ in model C) would lead to an estimate of 
$(\Omega_m, w_0) \approx (0.65, -0.6)$ if the data analysis was carried 
out with model B. These results mean that the same data can lead to different 
estimates of $\Omega_m$ and $w_0$, depending on the cosmological model 
used for the data analysis. 
In other words, the role of the derivative of the dark energy EOS on the
luminosity distance can be played by a matter density parameter in models with 
a constant EOS\,: a model with a fixed $\Omega_m$ containing a fluid described 
by a negative (positive) $w_1$ can be mistaken with a model 
described by a smaller (larger) $\Omega_m$ containing a barotropic fluid 
(see also Figure \ref {relc}). 
Note that the dotted curve corresponding to $z = 3$ shows 
degeneracy breaking at very large redshifts (see also Figure \ref {magnfig}).

\section{Quintessence models and present SNeIa data}
\label{present}

We used the published data of Perlmutter et~al. (1999; P99 hereafter), which 
consist of a sample containing 60 SNeIa, to determine the constraints that 
we have today on the parameters describing the four models introduced in 
Section~2. As it has been noted in P99, four of these supernovae are 
``outliers''. So we excluded them and 
considered a sample of only 56 SNeIa\footnote{Note that using the sample of 
60 SNeIa does not significantly change the results: the contour positions 
are approximately the same and the corresponding $\,\chi^2\,$ per degree of 
freedom is slightly larger.}. We analysed these data with a standard $\chi^2$ 
method, i.e. by minimizing the value of $\chi^2$ defined by
\begin{equation}
\displaystyle
\chi^2 = \sum_{i = 1}^{56}{\frac{\left[m^{th}_B (z_i; \theta_1, \theta_2,
\bar{M}_B) - m_{B,i}^{data}\right]^2}{\sigma_i^2}},
\label{chi2}
\end{equation}
where $m^{th}_B$ (resp. $m^{data}_B$) are the effective apparent magnitudes (i. e. 
the observed magnitudes corrected by the lightcurve width-luminosity relation, the 
Galactic extinction, the K-correction and the cross-filter calibration) predicted
by the model (resp. given by the data). $\sigma_i$ are the uncertainties on the 
data and $\theta_1$ et $\theta_2$ are the two parameters of the theoretical model 
(cf. Table \ref {tablemodel}). 

%
\begin{figure}
\begin{center}
\begin{tabular}{cc}
\includegraphics[height=10.2pc,width=9pc]{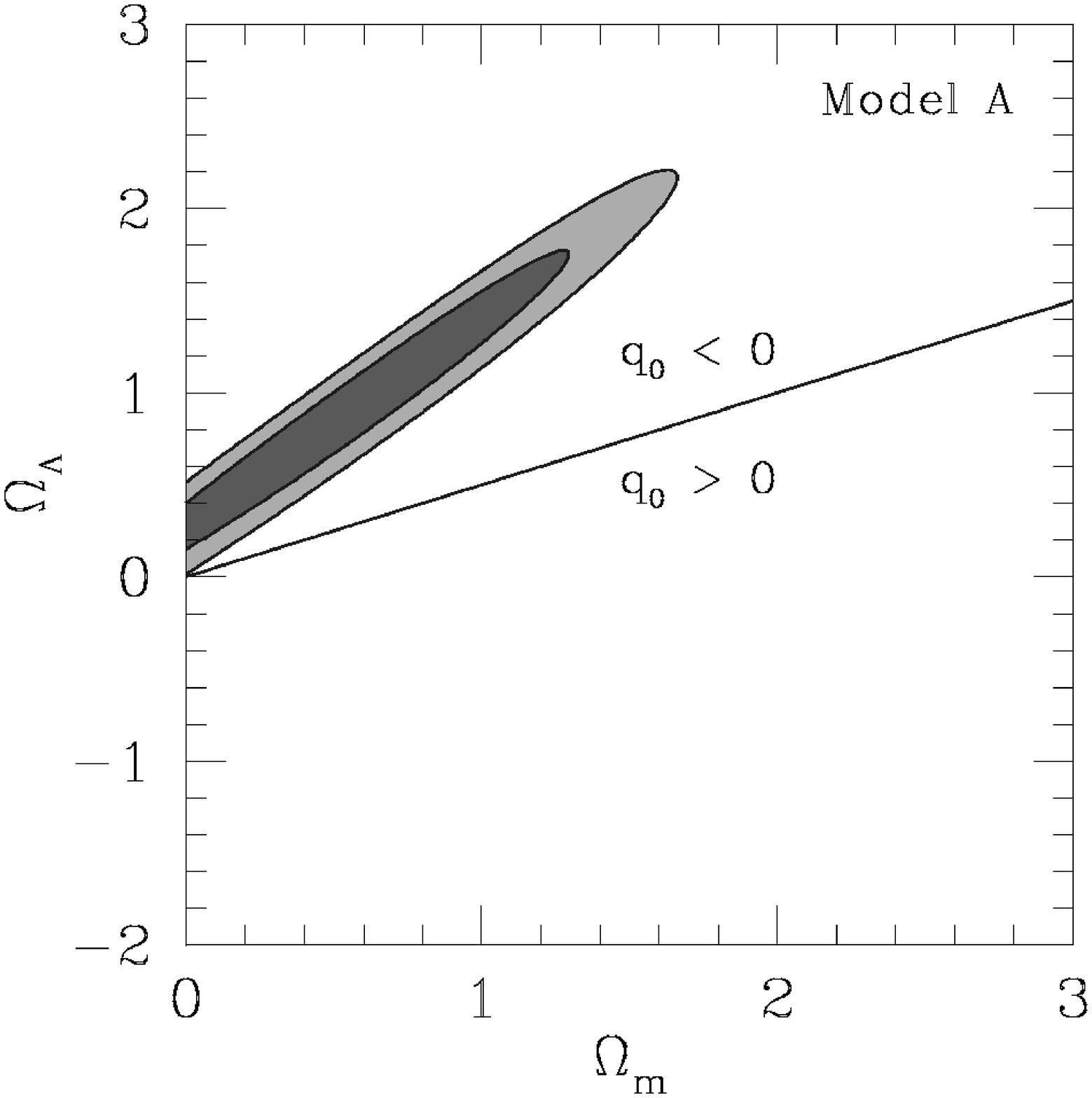} & 
\includegraphics[height=10.2pc,width=9pc]{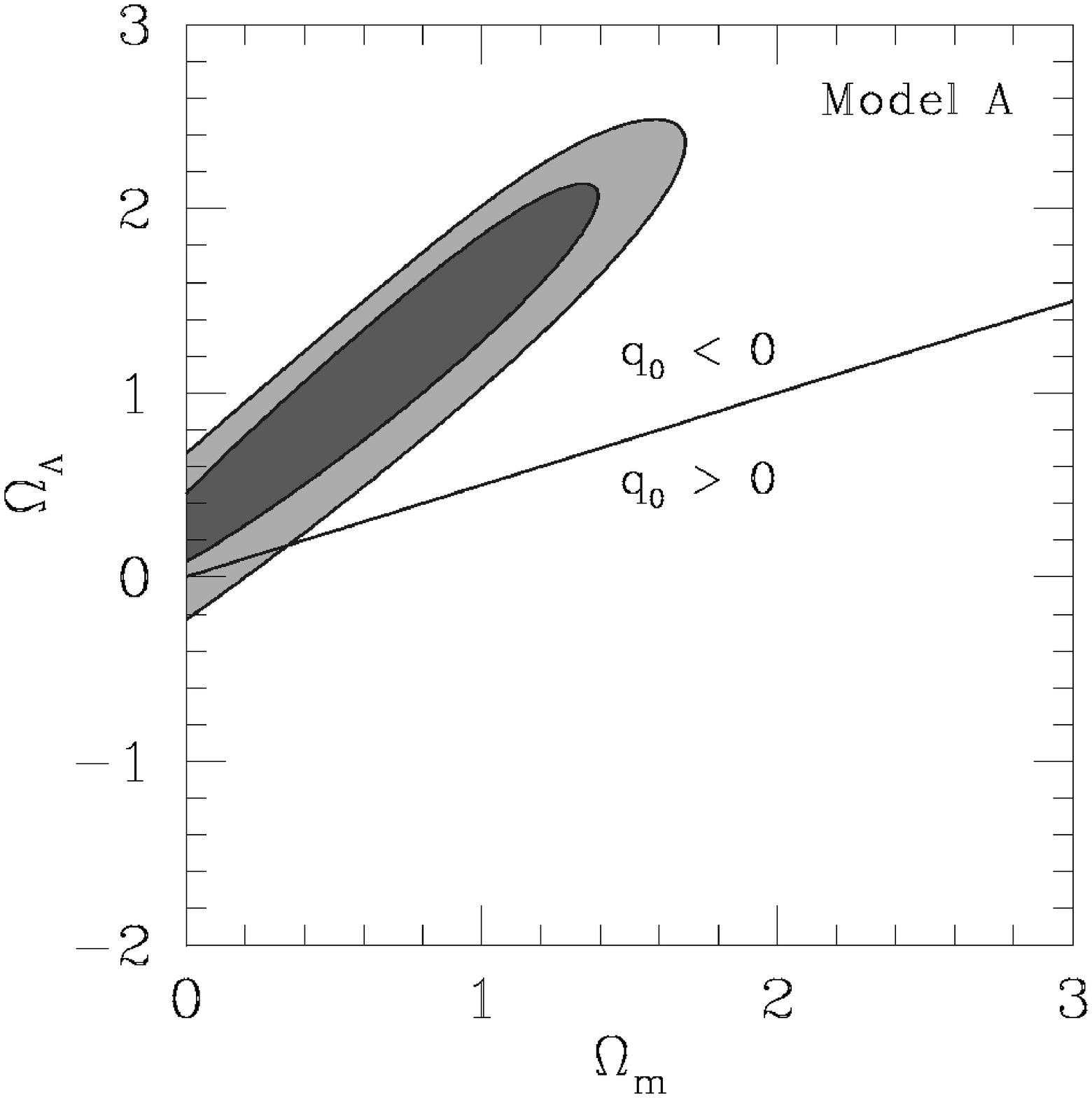} \\
\includegraphics[height=10.2pc,width=9pc]{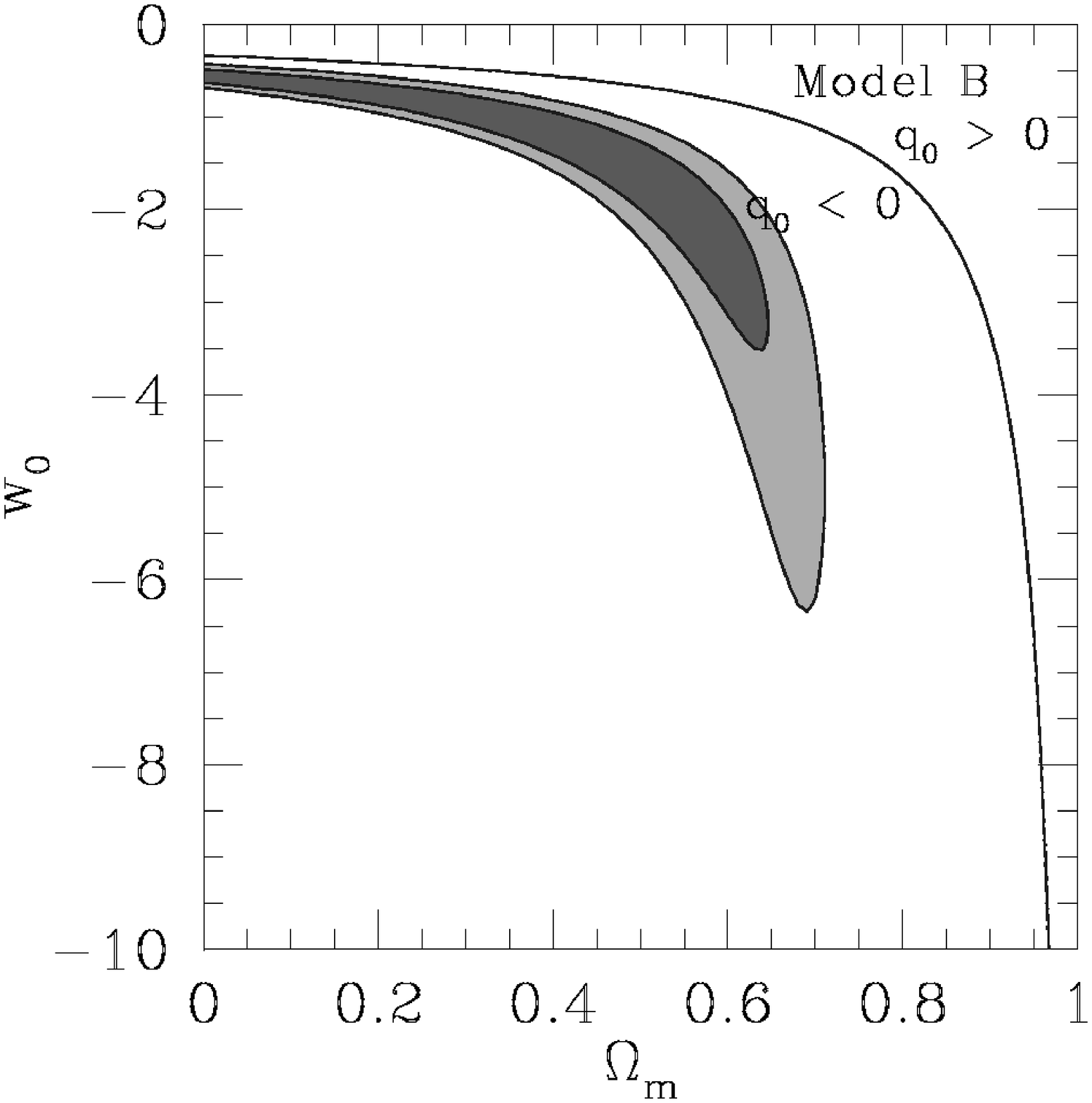} & 
\includegraphics[height=10.2pc,width=9pc]{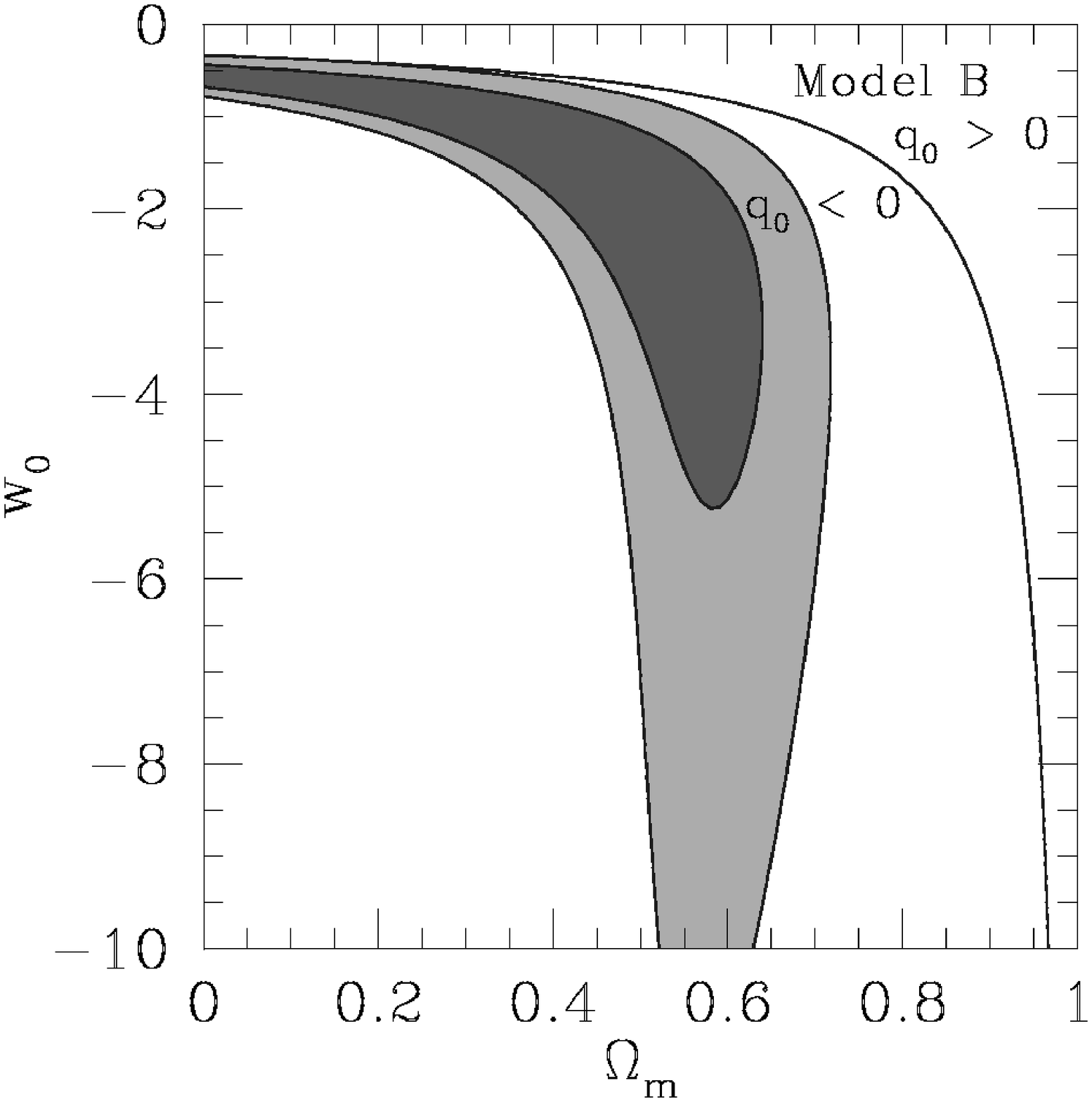} \\
\includegraphics[height=10.2pc,width=9pc]{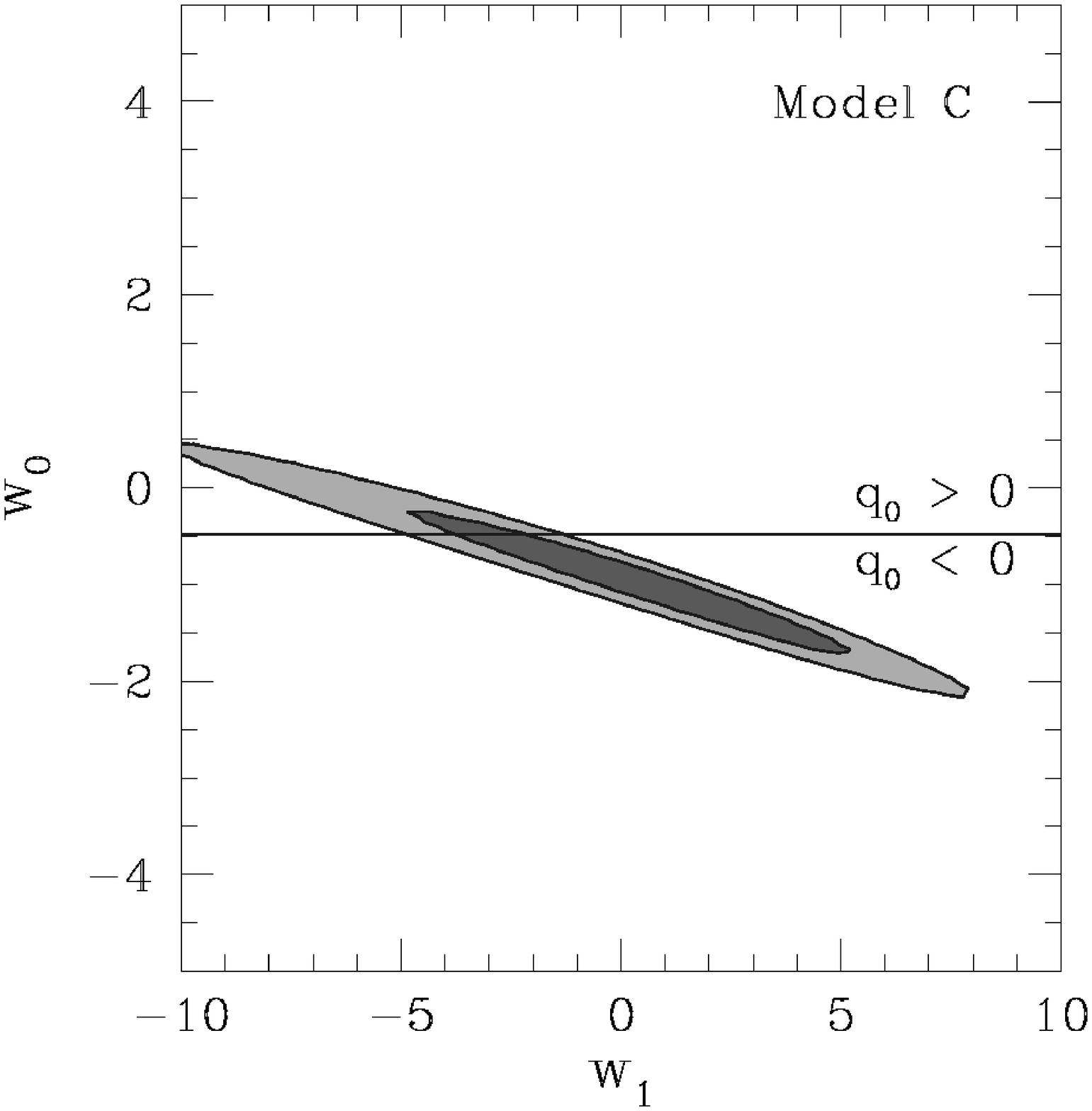} & 
\includegraphics[height=10.2pc,width=9pc]{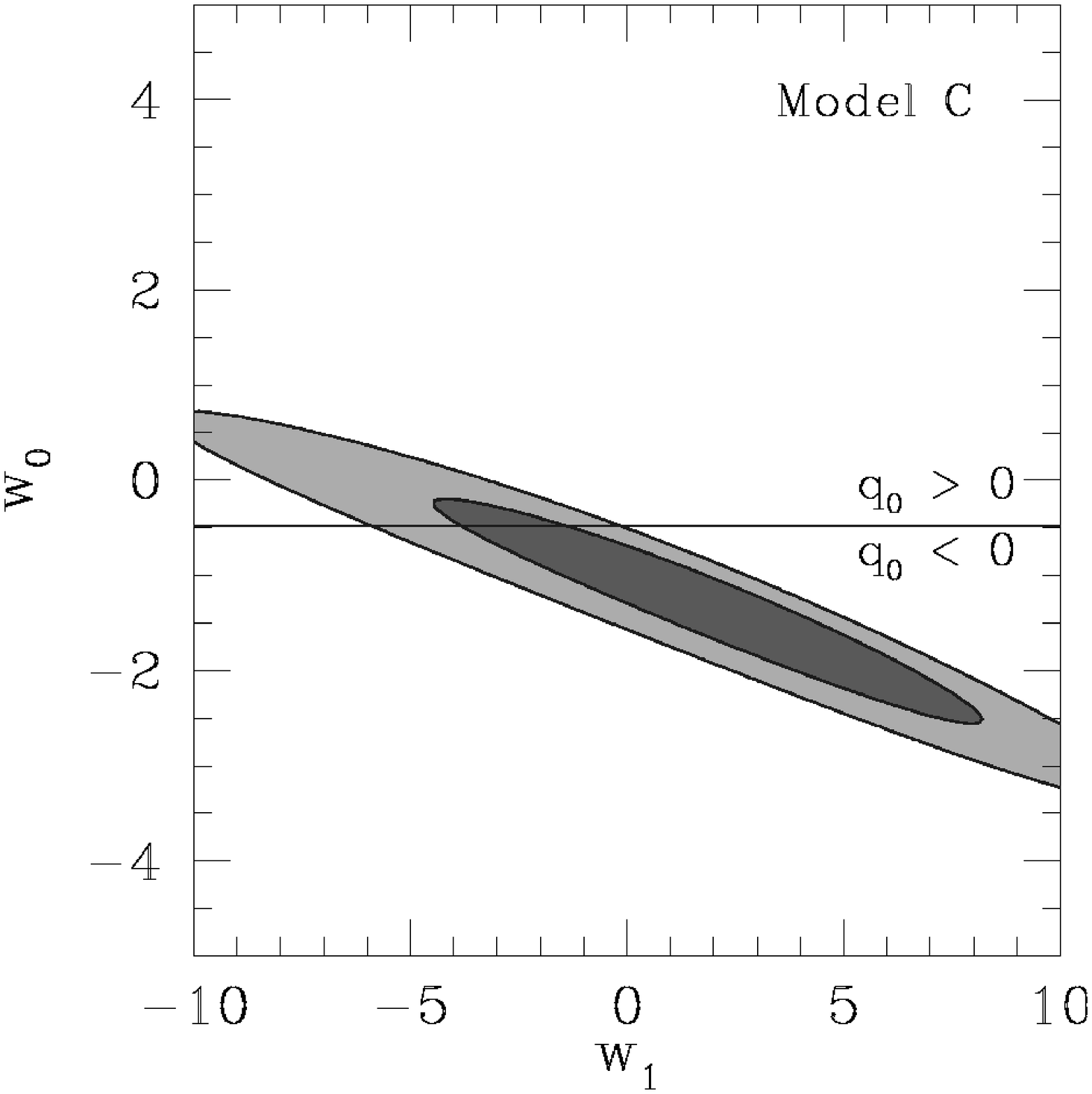} \\
\includegraphics[height=10.2pc,width=9pc]{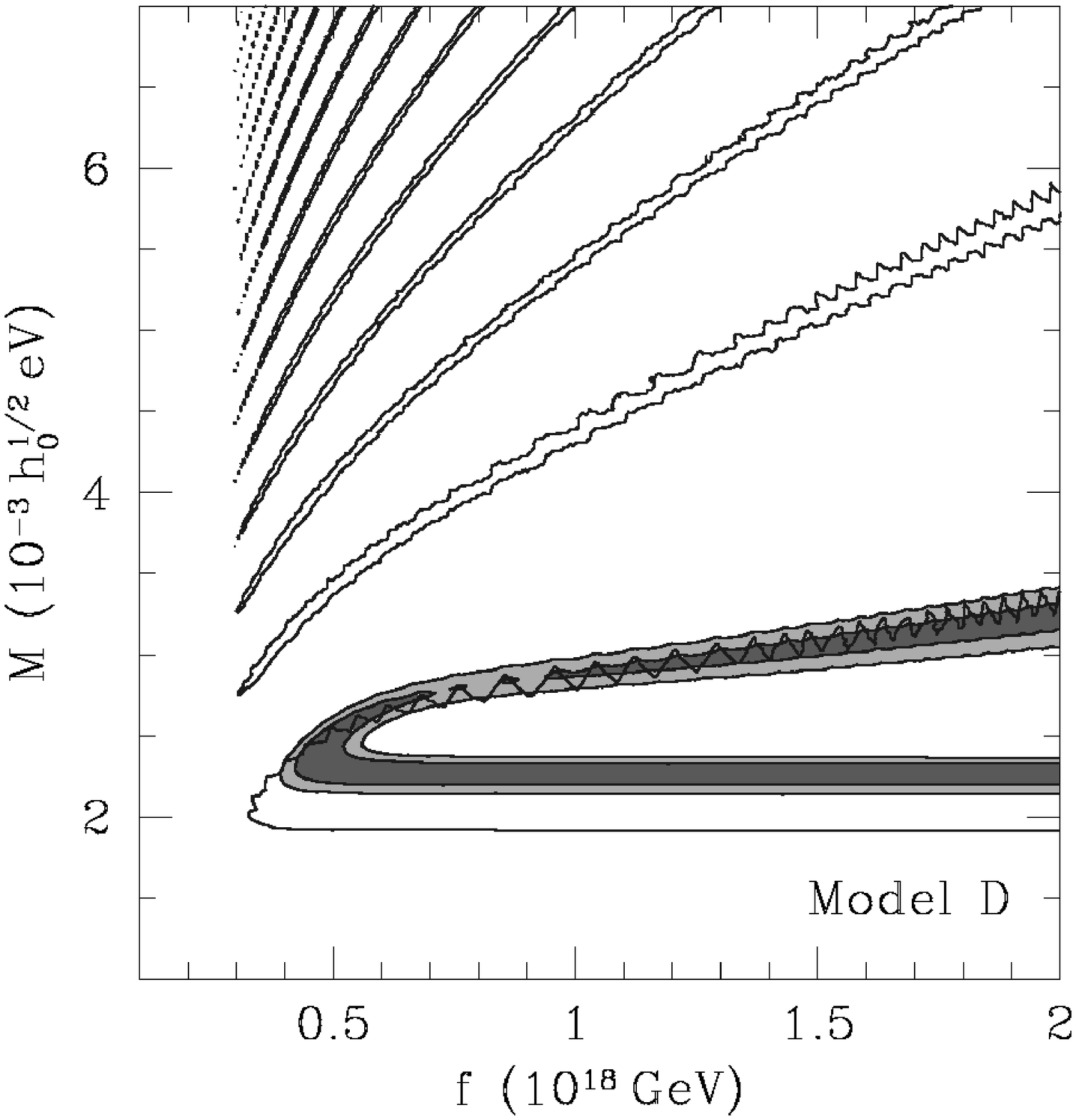} & 
\includegraphics[height=10.2pc,width=9pc]{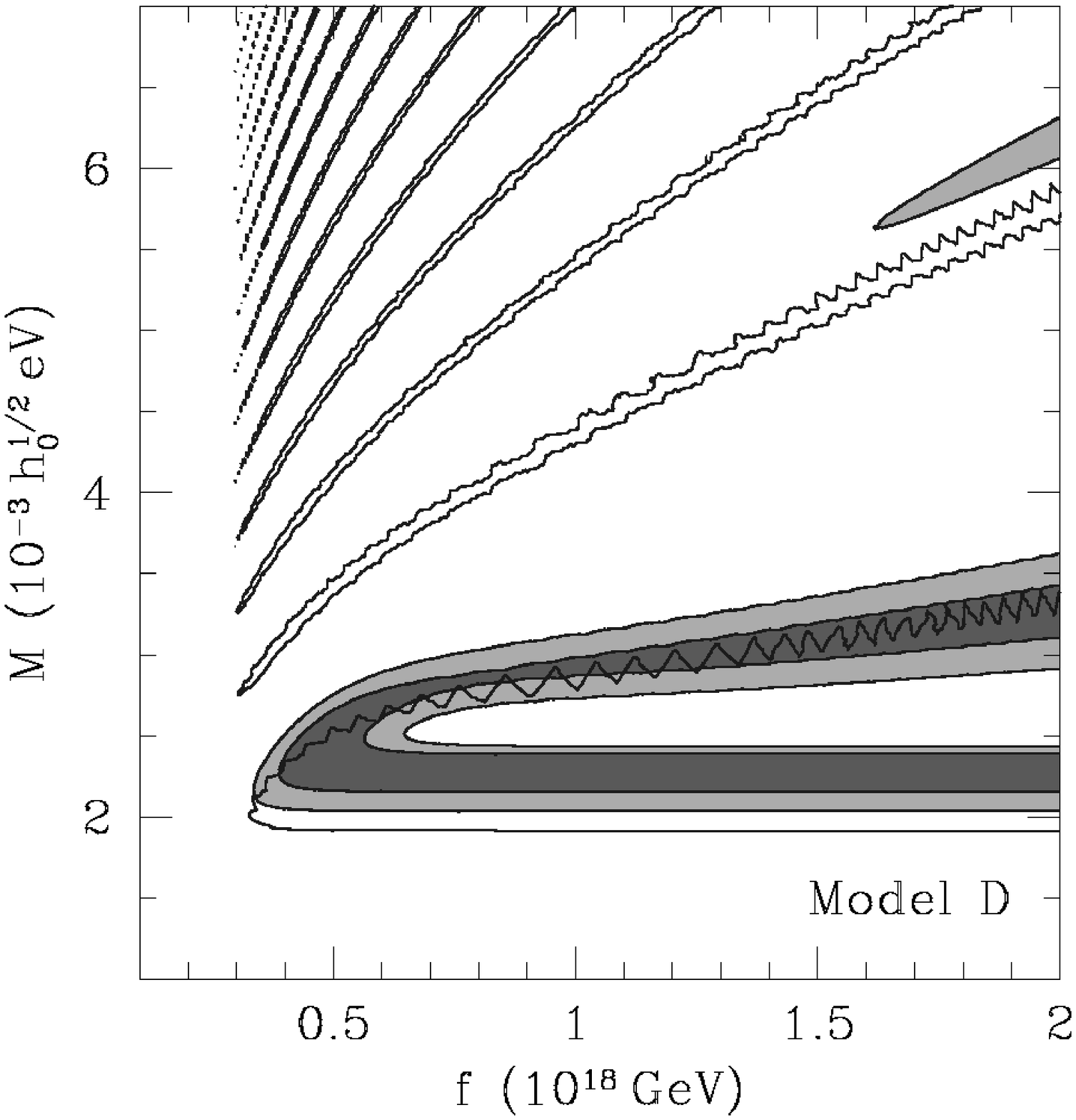}
\end{tabular}
 \caption{$68.3\%$ (dark) and $95.4\%$ (light) confidence contours obtained 
by analysing P99 data with each model. Solid curves separate universes which 
are presently decelerating from those which are presently accelerating. The 
four plots on the left (resp. on the right) are obtained with $\bar{M}_B = 
24$ (resp. with no prior constraints on $\bar{M}_B$).}
\label{figperlmutter}
\end{center}
\end{figure}

Concerning $\bar{M}_B$, we adopted two different approaches. We first
assumed that $\bar{M}_B$ was exactly known and equal to $24$,\footnote{This
value is the one obtained when $M_B = -19.5$ mag and $H_0 = 58.5$ km/s/Mpc are
introduced in relation (\ref {barm}) (Parodi et al. 2000).}
all uncertainties on this parameter being neglected. In the second approach,
we assumed no prior constraint on $\bar{M}_B$, which is just an unknown 
constant with a
value between $-\infty$ and $+\infty$. In this case, we integrated the
probabilities on $\bar{M}_B$ and therefore worked with a $\bar{\chi}^2$
defined by
\begin{equation}
\displaystyle
\bar{\chi}^2 = - 2\, \ln \left(
\int^{+\infty}_{-\infty}{e^{-\chi^2 / 2}\,d\bar{M}_B}
\right) = A - \frac{B^2}{C} + \ln \left( \frac{C}{2 \pi} \right),
\label{tildechi2}
\end{equation}
where
\begin{eqnarray}
A & = & \sum_{i = 1}^{56}{\frac{\left( m_{B,i}^{data} -
5\,\log[D_L^{th}]\right)^2} {\sigma_i^2}}, \\
B & = & \displaystyle
\sum_{i = 1}^{56}{\frac{m_{B,i}^{data} - 5\,\log[D_L^{th}]}{\sigma_i^2}}, \\
C & = & \displaystyle \sum_{i = 1}^{56} {\frac{1}{\sigma_i^2}}.
\end{eqnarray}
Note that a uniform distribution of the probabilities on a relatively small
interval, i.e. $\bar{M}_B = 24 \pm 0.2$, gives the same results as a uniform
distribution between $-\infty$ and $+\infty$. 

We analysed the sample of 56 SNeIa given in P99 using as framework the
four theoretical models discussed in Section~2 and for two opposite 
assumptions on $\bar{M}_B$. The probability contours are shown in Figure \ref 
{figperlmutter}. Table~\ref {tablemin2} lists the best fit values of the
parameters together with the $\chi^2_{\rm min}$ and  $\chi^2_{\rm min}/$DOF
(DOF\,: Degree of freedom).

The results obtained using model A are in agreement with those found in the 
literature (P99). Small differences may be related to the fact 
that we used data already corrected for the stretch-factor whereas P99
integrate on this parameter. However, these differences are 
not relevant in view of our objectives. Results concerning model B show that present
data agree with a baroropic fluid characterized by $w_0 < - 0.4$ at the $2\,\sigma$ 
level. An important remark has 
to be made concerning models B and C\,: Figure~\ref {figperlmutter} indicates that the
data favour unphysical regions where $w_0 < - 1$ (see also Caldwell 2002). Note that 
if we restrict ourselves to models with $w_0 \geq - 1$, the best fit positions, 
characterized by $(\Omega_m, w_0) = (0.3, -1)$ in model B and by $(w_0, w_1) = (-1, 
0.6)$ in model C, are associated to $\chi^2_{\rm min}$ values only marginally different 
from those obtained when $w_0 < -1$ ($\Delta\chi^2_{\rm min} < 0.1$). The fact that the 
true minimum is located in a unphysical region is something that has to be strongly 
considered in the future because, if confirmed, it can rule out these models. 

\begin{table}
\begin{center}
\begin{tabular}{|c||c|c|c|c|}
\hline
Model & $\chi^2_{\rm min}$ & $\chi^2_{\rm min}/$DOF & $\theta_1$ & $\theta_2$  \\
\hline
$A$ & $62.1$ & $1.15$ & $\Omega_m = 0.6$ & $\Omega_\Lambda = 0.9$ \\
$B$ & $62.2$ & $1.15$ & $\Omega_m = 0.41$ & $w_0 = - 1.2$  \\
$C$ & $62.2$ & $1.15$ & $w_0 = - 1.1$ & $w_1 = 0.9$ \\
$D$ & $62.2$ & $1.15$ & $M = 2.3$ & $f = 0.5$ \\
\hline
\end{tabular} \\ 
\begin{tabular}{|c|c|c|c|c|}
\hline
Model & $\bar{\chi}^2_{\rm min}$ & $\bar{\chi}^2_{\rm min}/$DOF & $\theta_1$ & $\theta_2$  
\\ 
\hline
$A$ & $66.8$ & $1.26$ & $\Omega_m = 0.8$ & $\Omega_\Lambda = 1.3$ \\
$B$ & $67.4$ & $1.27$ & $\Omega_m = 0.46$ & $w_0 = - 1.6$ \\
$C$ & $67.3$ & $1.27$ & $w_0 = - 1.4$ & $w_1 = 2.3$\\
$D$ & $67.4$ & $1.27$ & $M = 2.3$ & $f = 0.5$ \\
\hline
\end{tabular}
\caption{Minimum values of $\chi^2$ and $\bar{\chi}^2$ obtained when the analysis 
of the existing data is carried out on each of the four quintessence models. The 
upper table assumes $\bar{M}_B = 24$ whereas the lower one assumes no prior 
constraints on $\bar{M}_B$.}
\label{tablemin2}
\end{center}
\end{table}

The constraints for model D select two sorts of regions (Figure \ref 
{figperlmutter}\,; see also Ng \& Wiltshire 2001b)\,: 
one where the scalar field is still frozen (lower horizontal band) and another where 
its evolution has just begun (upper oblique band). Note that these contours are not 
closed\,: the larger the $f$, the more the two branches are separated. A small part 
of the $2\,\sigma$ contours also appears on the second plot of model D, selecting a 
region with $M \sim 6 \times 10^{-3}\,h_0^{1/2}$ eV. If we adopt the point-of-view 
that the PNGB field of model D is related to the resolution of the solar neutrino 
problem, then the value of $M$ gives an estimate of the neutrino mass 
(Frieman et~al. 1992, 1995). The last two 
plots of Figure~\ref {figperlmutter} show that at least one neutrino species has a mass 
of\,: $\sim 2 \times 10^{-3}\,h_0^{1/2}$ eV, if the PNGB field is still frozen, $\sim 
3 \times 10^{-3}\,h_0^{1/2}$ eV, if the field is just starting to evolve or $\sim 6 
\times 10^{-3}\,h_0^{1/2}$ eV ($2\,\sigma$ region on the second plot), if the field has 
already passed one time through the potential minimum. These estimates are indeed 
similar to the ones found by SuperKamiokande and SNO (Sudbury Neutrino Observatory) 
for the neutrino mass squared 
differences, i.e. $\delta m_{23}^2 = (1.6 - 3.9) \times 10^{-3}$ eV$^2$ (Fukuda et al. 
1998) and $\delta m_{12}^2 = (0.3 - 3.5) \times 10^{-4}$ eV$^2$ (Choubey et al. 2002 
and references therein). This may be only a numerical coincidence... or it may comfort
us in the resolution of the neutrino problem through a Goldstone mechanism. 

Note that contours where $M \sim 3 \times 10^{-3}\,
h_0^{1/2}$ eV or $6 \times 10^{-3}\,h_0^{1/2}$ eV are associated to small values of 
$\Omega_m$ ($\ll 0.1$\,; see Figure~\ref {figclassifd}) which are inconsistent with 
observations (Peebles \& Ratra 2002). The value of $f$ is only poorly constrained 
by the present data. This 
is because $f$ becomes an important parameter for the cosmological model only when the 
scalar field evolves in its potential, not when it is frozen or just starting to move.  

As it can be seen on Figure \ref {figperlmutter}, in some cases, present data can also 
favour a decelerating universe, even at $1\,\sigma$ level.
An important remark has to be made concerning the $\chi^2_{\rm min}$ values presented
in Table~\ref {tablemin2}\,: in both approaches for $\bar{M}_B$, the four values 
of $\chi^2_{\rm min}$ are very close to each other. In particular, the differences 
between them are much less than the $1\,\sigma$ interval ($\Delta \chi^2= 2.3$).
Therefore, as expected, the present data are not sufficient to discriminate
between these four quintessence models even at the $1\,\sigma$ confidence level.

\section{Quintessence models and future SNeIa observational data}
\label{snap}

The analysis of the present SNeIa data discussed in the previous section
clearly showed that more observations are needed to clearly determine the 
nature of the dark energy. Such data could be provided by the proposed 
SNAP (``SuperNova/Acceleration Probe'') mission (see SNAP URL). SNAP, if 
accepted, will be a two-meter space telescope dedicated to the discovery 
and observation of SNeIa with redshifts in the range $0 < z < 1.7$. This 
instrument should perform photometry and spectroscopy of more than 
2000 SNeIa per year. The aim of this section is to investigate which values
of the parameters (chosen by Mother Nature), future SNAP observations will be able
to discriminate. The analysis will also reveal degeneracies between individual
parameters of the different models.

We have first generated mock SNAP data by successively assuming that model A, B, 
C or D was the true one, and under the ideal
hypothesis that $\bar{M}_B$ was exactly known\footnote{Other works directly dealing 
with luminosity distance reconstruction instead of magnitude reconstruction assume 
that $\bar{M}_B$ is perfectly known (Huterer \& Turner 1999, Saini et al. 2000, 
Chiba \& Nakamura 2000, Gerke \& Efstathiou 2002).}. For each model, we have 
selected 100 couples of parameters restricted to 
physical regions surrounding contours given on Figure~\ref {figperlmutter} (see 
Table~\ref {samples}). This is why the $w_0$ parameter has been limited to $[-1; 
0]$. The $w_1$ parameter of model C has been limited to the interval 
$[-1; 1]$, in order to avoid large absolute values of the EOS at $z \geq 1$. 
The parametrization adopted to generate data from 
model D means that we restricted ourselves to universes different from an 
Einstein-de Sitter one 
($M < 5$) or to universes characterized by $\Omega_m < 0.6$ ($f > 0.4$ or $M > 2$) (cf. 
Figure~\ref {figclassifd}).

\begin{figure}
\begin{center}
\includegraphics[height=28pc]{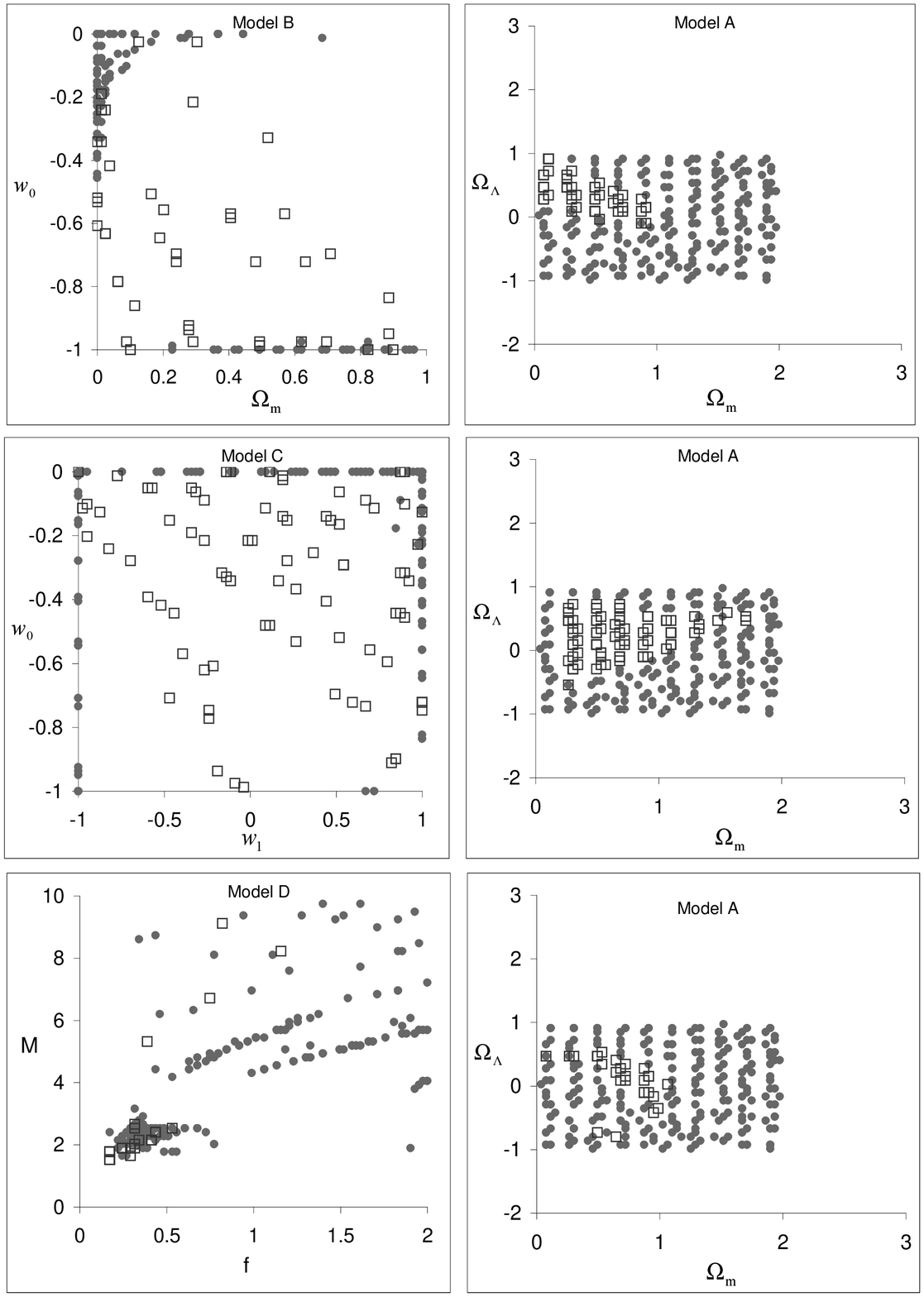} 
\caption{
Positions of the $\chi^2_{\rm min}$ in the four models of 
the $300$ simulated data samples generated with model A.
The filled circles (resp. empty squares) correspond to 
$\Delta\,\chi^2_{\rm min,X-A} > 6.17$ (resp. 
$\Delta\,\chi^2_{\rm min,X-A} < 6.17$) for $X = B,C,D$.}
\label{figdota}
\end{center}
\end{figure}

Then, for each value of the parameters, we have generated 
three samples. For each of them we have computed the magnitude of 2100 SNeIa, 
with 2000 uniformly distributed in the redshift interval $z \in [0 ; 1.2]$ 
and 100 in the redshift interval $z \in [1.2 ; 1.7]$. Gaussian errors on the 
magnitude have been added with a standard deviation of 0.15 mag, reflecting only 
the intrinsic spread of the SNeIa maximum luminosity (see SNAP URL). So for each 
of the four models, we are left with $3 \times 100$ sets of SNAP data, for which 
we know the parent model. Then, each sample has been independently fitted with the 
four models. The results are presented in Figures \ref {figdota}, \ref{figdotb}, 
\ref{figdotc} and 
\ref {figdotd} where the parent is model A, B, C, D respectively. The parent model
is in the right column and is successively compared to the other models. The squared 
symbols select the value of the parameters for which the two models are degenerated 
at the $2\,\sigma$ level, i.e. for which $\Delta\,\chi^2_{\rm min} < 6.17$, whereas 
the filled 
dots represent the parameters for which $\Delta\,\chi^2_{\rm min} > 6.17$ . Since we 
simulated noisy data, fitting the true model may not yield the exact value of the 
parameters. That is why the sets of models do not reproduce a regular grid in the
right column of Figures \ref {figdota}, \ref {figdotb}, \ref{figdotc} and \ref 
{figdotd}. Moreover, different best fit positions can be associated to several 
simulations. 

\begin{table}
\begin{center}
\begin{tabular}{cl}
Model & Values used for generate samples \\
\hline
A & $\Omega_{m,i} = 0.1 - (1 - i)\,/\,5$ \\ 
& $\Omega_{\Lambda,j} = -0.9\, + 0.2 \,(j - 1)$ \\ & \\
B & $\Omega_{m,i} = 0.05 + (i - 1)\,/\,10$  \\
& $w_{0,j} = - 0.05 - (j - 1)\,/\,10$  \\ & \\
C & $w_{0,i} = -0.05 + (1 - i)\,/\,10$ \\
& $w_{1,j} = -0.95 \,(2\,j - 11)\,/\,9$ \\ & \\
D & $f_i = 0.4 + 1.6\,(i - 1)\,/\,9\,$ \\ 
& $M_j = 2 + (j - 1)\,/\,3\,$  \\
\hline
\end{tabular}
\caption{Values of the parameters used for the SNAP data simulations 
($i,j = 1, ..., 10$).}
\label{samples}
\end{center}
\end{table}

\begin{figure}
\begin{center}
\begin{tabular}{c}
\includegraphics[height=14pc,width=18pc]{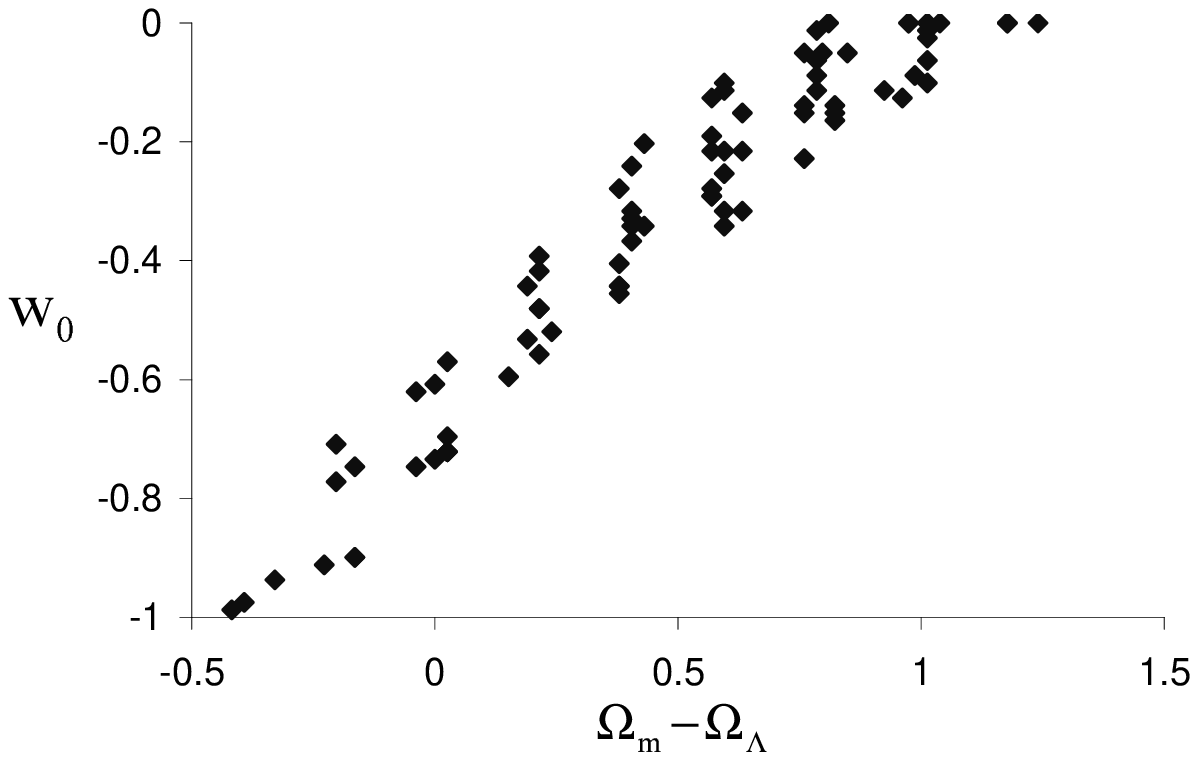} \\
\includegraphics[height=14pc,width=18pc]{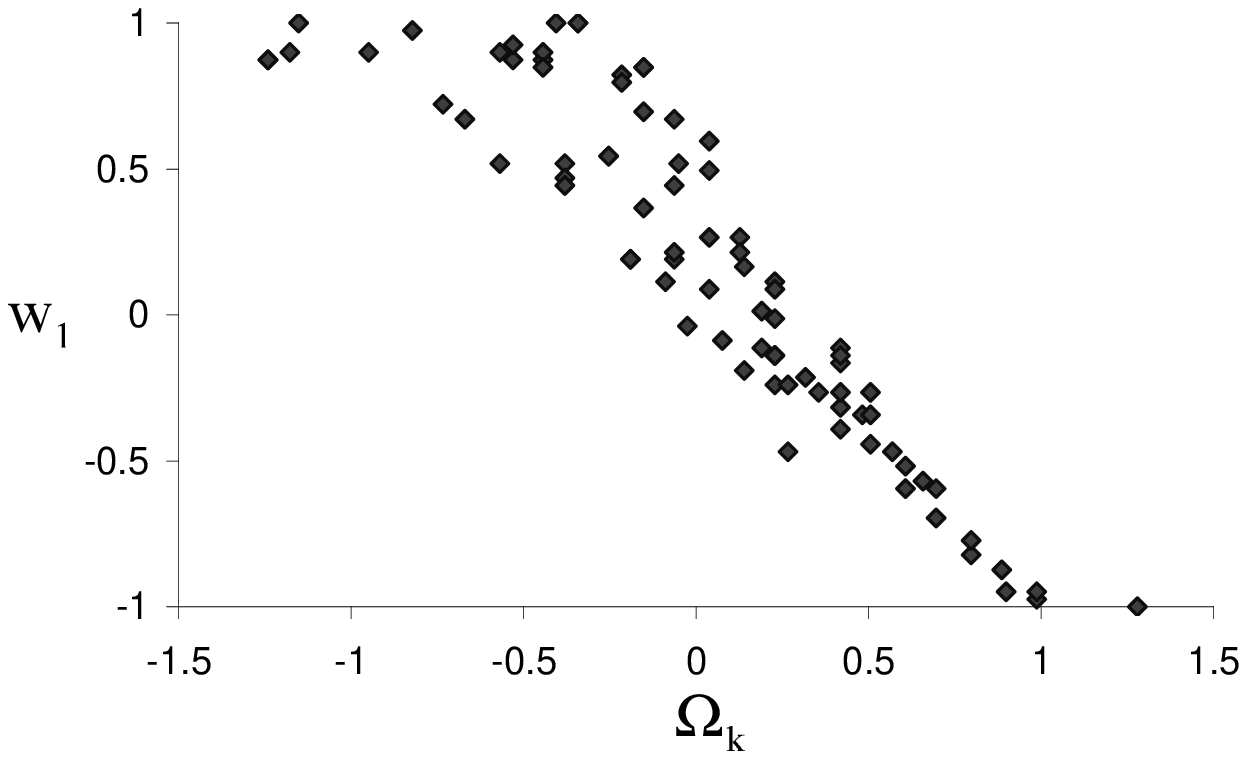} 
\end{tabular}
\caption{Correlations between individual parameters of model A and
model C (when $\Delta \chi^2_{\rm min} < 6.17$). {\em Upper plot - } 
between $\Omega_m - \Omega_\Lambda$ of the correct model A and $w_0$ 
of the wrong model C, {\em Lower plot - } between $\Omega_k$ of the 
correct model A and $w_1$ of the wrong model C.}
\label{rela}
\end{center}
\end{figure}

\begin{figure}
\begin{center}
\includegraphics[height=28pc]{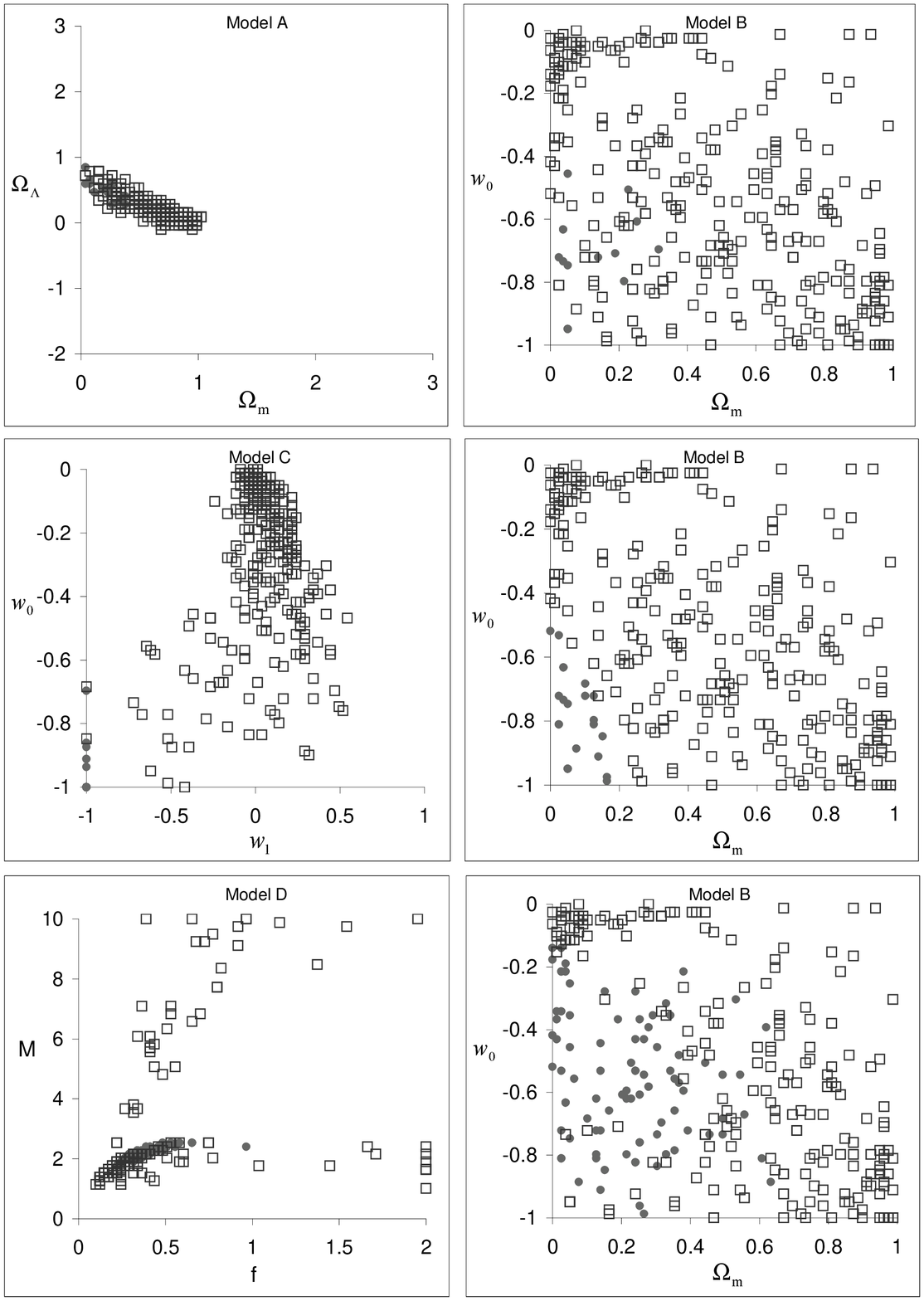} 
 \caption{
Positions of the $\chi^2_{\rm min}$ in the four models of the $300$ 
simulated data samples generated with model B.
The filled circles (resp. empty squares) correspond to $\Delta\,
\chi^2_{\rm min,X-B} > 6.17$ (resp. $\Delta\,\chi^2_{\rm min,X-B} 
< 6.17$) for $X = A,C,D$.}
\label{figdotb}
\end{center}
\end{figure}

\begin{figure}
\begin{center}
\begin{tabular}{c}
\includegraphics[height=14pc,width=18pc]{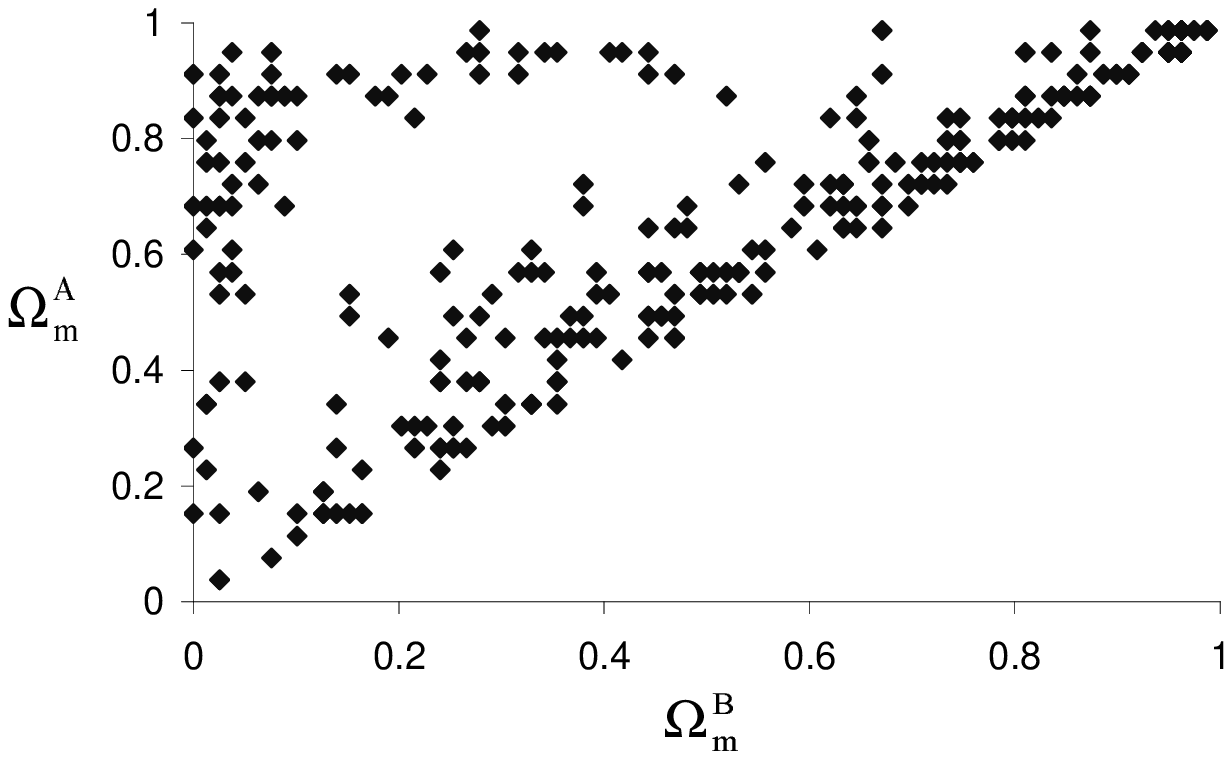} \\
\includegraphics[height=14pc,width=18pc]{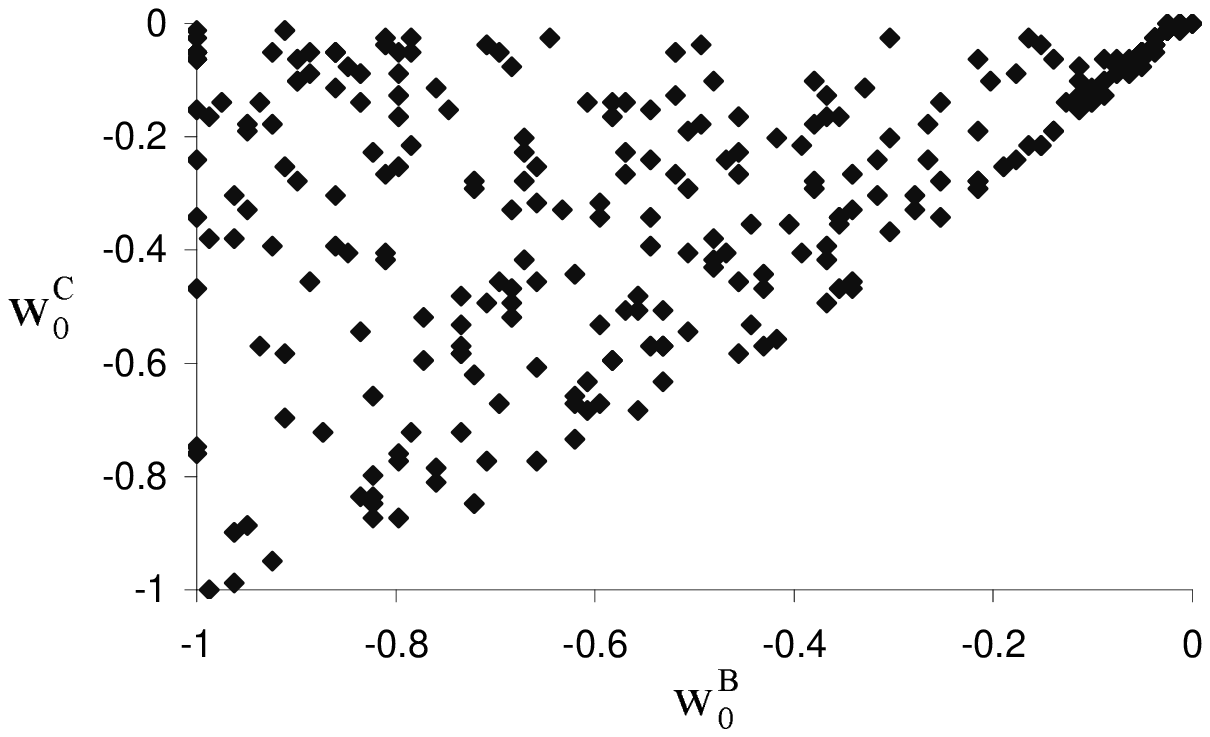} 
\end{tabular}
\caption{{\em Upper plot - } Correlation between the best fit 
values of $\Omega_m$ obtained with the correct model B and 
with the wrong model A, when $\Delta \chi^2_{\rm min} < 6.17$. 
{\em Lower plot - } Correlation between the best fit values of $w_0$
obtained with the correct model B and with the wrong model C, 
when $\Delta \chi^2_{\rm min} < 6.17$.}
\label{relb}
\end{center}
\end{figure}

\begin{figure}
\begin{center}
\includegraphics[height=28pc]{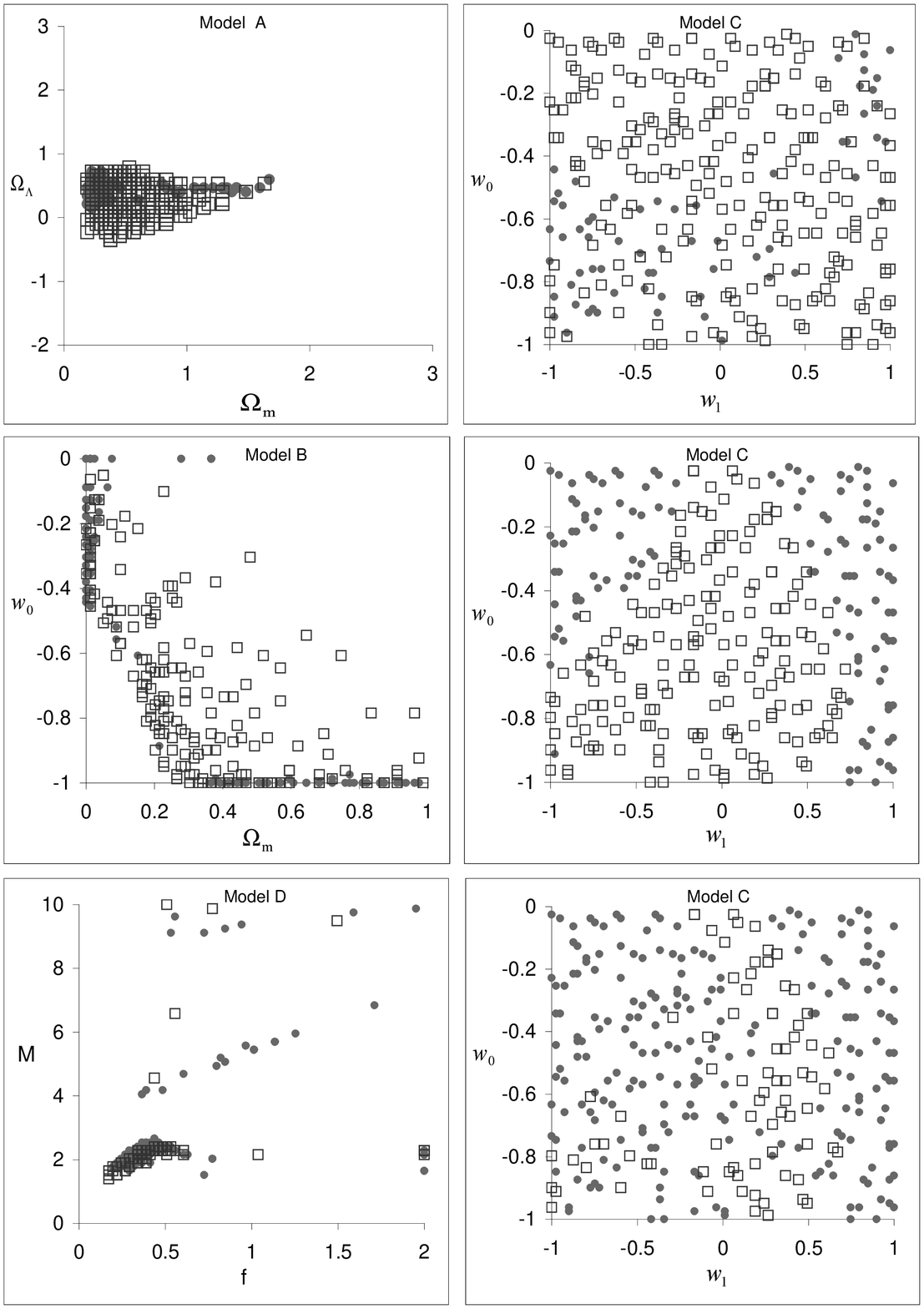} 
 \caption{
Positions of the $\chi^2_{\rm min}$ in the four models of the $300$ 
simulated data samples generated with model C.
The filled circles (resp. empty squares) correspond to 
$\Delta\,\chi^2_{\rm min,X-C} > 6.17$ (resp. $\Delta\,
\chi^2_{\rm min,X-C} < 6.17$) for $X = A,B,D$.}
\label{figdotc}
\end{center}
\end{figure}

\begin{figure}
\begin{center}
\includegraphics[height=14pc,width=18pc]{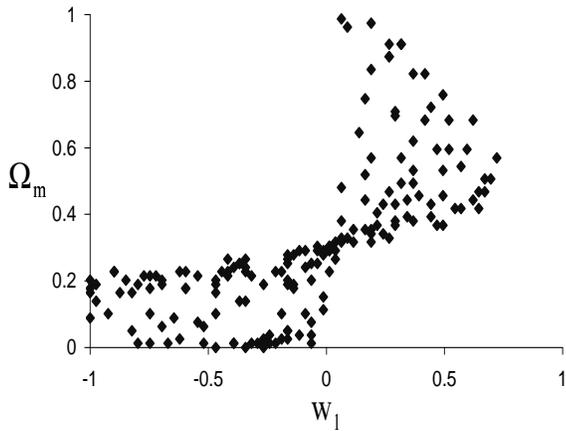} 
\caption{Correlation between the best fit values of $w_1$
obtained with the correct model C and the best fit values 
of $\Omega_m$ obtained with the wrong model B, when 
$\Delta \chi^2_{\rm min} < 6.17$.}
\label{relc}
\end{center}
\end{figure}

\begin{figure}
\begin{center}
\includegraphics[height=28pc]{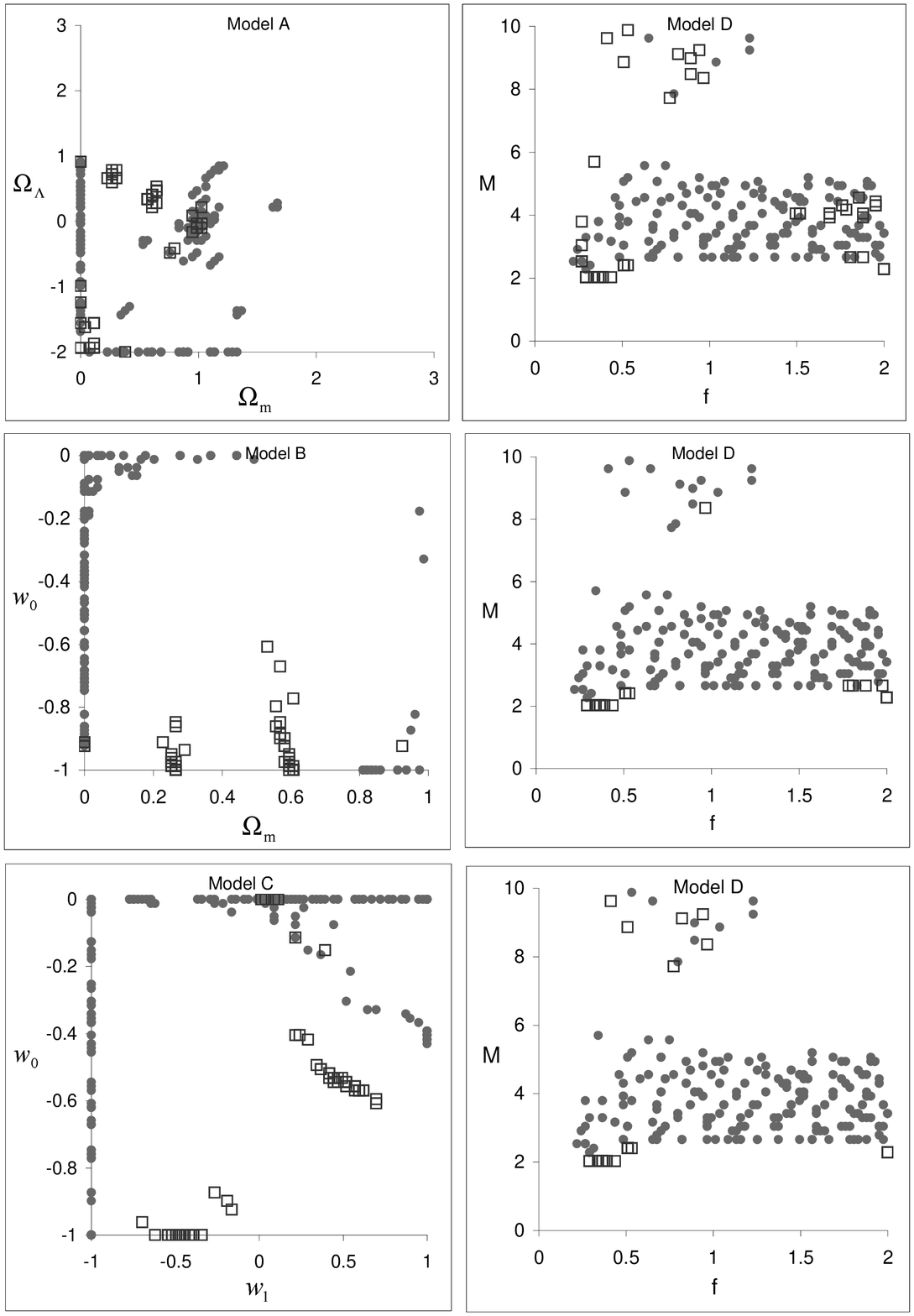} 
\caption{
Positions of the $\chi^2_{\rm min}$ in the four models of the $300$ 
simulated data samples generated with model D.
The filled circles (resp. empty squares) correspond to 
$\Delta\,\chi^2_{\rm min,X-D} > 6.17$ (resp. $\Delta\,\chi^2_{\rm min,X-D} 
< 6.17$) for $X = A,B,C$.}
\label{figdotd}
\end{center}
\end{figure}

\subsection{Analysis of data generated with model A}
Figure \ref {figdota} shows results obtained by analysing data generated from
model A. It shows that it could be difficult to 
unambiguously distinguish a model A with $\Omega_k = 0$ from a model B, C or 
D at $2\,\sigma$ level. Some misleading results can also be obtained from a
model A with $\Omega_k \not= 0$. This displays degeneracies between 
the curvature and the EOS of dark energy\,: a FLRW model with spatial curvature
($\Omega_k \not= 0$) and a cosmological constant ($w(z) = - 1$) can be confused 
with a spatially flat model containing a dark
fluid with a more general EOS (see also Figure \ref {rela} below). 
On the other hand, points located on the edges
of the considered parameter spaces correspond to model A values for which 
it is unlikely to obtain a good fit with a wrong model, i.e. for 
which reliable results can be expected. Let us also mention the following 
points\,:\\ \noindent
$-$ With model B, misleading results are
obtained if the values of $(\Omega_m, \Omega_\Lambda)$ are confined in the 
region inside the two lines $\Omega_k = 0$ and $\Omega_\Lambda = 0$. 
\\ \noindent
$-$ Ambiguous results between models A and C are expected 
if the true model is model A with $(\Omega_m, \Omega_\Lambda)$
satisfaying these three conditions\,: $\Omega_m + \Omega_\Lambda 
< 1$, $\Omega_m - \Omega_\Lambda < 1$ and $\Omega_m > 0.25$. \\ \noindent
$-$ In comparison with model D, most of the squares are located on the line 
$\Omega_k = 0$ with $\Omega_m > 0.5$ in the parameter space of model A. This 
means that there is no degeneracy between the EOS of a PNGB fluid and the 
curvature parameter. In addition, more than half of the $\chi^2_{\rm min,D}$ 
positions are located in a region where $M \sim 2$ and $f \sim 0.4$, i.e. 
characterized by $\Omega_m > \Omega_x$ and by a scalar field which is 
presently frozen or is just starting to evolve through the minimum of its 
potential (cf. Figure \ref {figclassifd}).

Finally, Figure \ref{rela} shows the correlations rising between individual 
parameters of model A and model C when both models are undistinguished at 
$2\,\sigma$ level ($\Delta \chi^2_{\rm min} < 6.17$). Linear regressions 
through the points yield\,:
\begin{eqnarray}
w_0 & \sim & \displaystyle \frac{2}{3}\,\left( \Omega_m - \Omega_\Lambda 
- 1 \right), \label{blabla1} \\
w_1 & \sim & \Omega_m + \Omega_\Lambda - 1 \,\,= - \Omega_k. \label{blabla2}
\end{eqnarray}
This demonstrates the degeneracy between the curvature of an FLRW
model and the $w_1$ parameter of a model C. Note that by construction, model C 
with $w_0 = - 1$ and $w_1 = 0$ is identical to an FLRW model with $\Omega_m = 0.3$ 
and $\Omega_\Lambda = 0.7$. This result is basically reproduced by the
relations (\ref {blabla1}) and (\ref {blabla2}).

\subsection{Analysis of data generated with model B}
Figure \ref {figdotb} shows the expected degeneracies\, if the true model is 
a model B. The irregurar distribution of the best fit positions in the parameter 
space of model B denotes its stronger sensitivity to the statistical noise. 
Moreover, since we looked at the $\chi^2_{\rm min}$ scanning parameter space 
from $\Omega_m = 0$ to $\Omega_m = 1$, we find an accumulation of points in the 
upper left corner of the parameter space of model B, as expected from the form 
of the degeneracy curves shown in Figure \ref {figdeg4}. Note that the first row 
of Figure \ref {figdotb} shows again the degeneracy between $\Omega_k$ and $w_0$, 
as already noted in Figure \ref {figdota}\,: a flat model B with $w_0 \not= - 1$ 
can be confused with an FLRW model with spatial curvature.

Figure \ref {relb} shows the possible error on the estimate of $\Omega_m$ and 
$w_0$, when the correct model B is not used to fit the data. Although there is no 
significant difference in terms of $\chi^2_{\rm min}$, 
fitting the data with an FLRW model
yields values for $\Omega_m$ larger than or equal to the exact value (upper plot),
and fitting the data with a model C often give rise to value of $w_0$ larger than 
or equal to the correct one (lower plot).

\subsection{Analysis of data generated with model C}
Figure \ref {figdotc} shows the best fit positions in the parameter spaces of 
models A, B and D when model C has been used as a 
fiducial model. One can see that\,:
\\ \noindent
$-$ First, the $\chi^2_{\rm min,A}$ are all located in a region charaterized 
by $\,\Omega_m + \Omega_\Lambda < 1$ and $\,\Omega_m - \Omega_\Lambda 
< 1$. This confirms what has been obtained in Figure \ref {figdota} (cf. 
second plot of model A on Figure \ref {figdota}). \\ \noindent
$-$ Second, as expected, the 
$\Delta\,\chi^2_{\rm min}$ between models B and C less than $6.17$ are usually 
associated to the model C with $w_1 \sim 0$, but only if $w_0 > - 0.4$. 
\\ \noindent
$-$ Third, as previously noted, most of the 
$\chi^2_{\rm min}$ positions in the parameter space of model D are located in a 
region where the scalar field is presently starting to evolve toward the minimum 
of its potential, i.e. $w_x(z)$ leaves the value $-1$ and approaches unity, 
with $\Omega_\phi > \Omega_m$. This is why in most cases, the corresponding 
squares in model C are characterized by $w_1 \geq 0$. 

Figure \ref {relc} shows the correlation between the best fit values of $w_1$
obtained with the true model (C) and the best fit values of $\Omega_m$ obtained 
with a wrong model (B), when $\Delta \chi^2_{\rm min} < 6.17$. If the true 
value of $w_1$ is positive (negative), then the value of $\Omega_m$ found 
with the wrong model will be larger (smaller) than the true one ($\Omega_m = 0.3$). 

%
\subsection{Analysis of data generated with model D}
Finally, Figure \ref {figdotd} presents results obtained using samples 
generated with model D. It clearly appears that most of the $\chi_{\rm min}^2$ 
positions corresponding to $\Delta\chi^2_{\rm min} > 6.17$ are located on the 
edges of the parameter spaces of the wrong models. So if the cosmological 
model chosen by Mother Nature is a PNGB model, it could be difficult 
to obtain a good fit with one of the three wrong models A, B or C. 
Moreover, the presence of some points outside the region of model D used 
for generating samples denotes a sensitivity to statistical noise. Note that 
most of the squares in the parameter space of model A of this figure appear 
like three bundles located at $\Omega_m \sim 1 - \Omega_\Lambda \sim 0.3, 0.6$
and $1$. This is due to the strong dependence of $\Omega_m$ in model D with $M$ 
when the scalar field is still frozen and acts like a cosmological constant (cf. 
horizontal lines for $\Omega_m$ in Figure \ref {figclassifd}). The more precise 
is the grid used for the data processing in model D, the more those squares will 
appear uniformly distributed on the line $\Omega_m + \Omega_\Lambda = 1$ in the 
parameter space of model A.

\begin{table*}
\begin{center}
\begin{tabular}{l|ccc|ccc|ccc|ccc}
\hline
True model & & A & & & B & & & C & & & D &  \\
 & \multicolumn{3}{|l|}{\small $\Omega_{m,i} =0.1-(1-i)/5$} & 
\multicolumn{3}{|l|}{\small $\Omega_{m,i}=0.05+(i-1)/10$} & 
\multicolumn{3}{|l|}{\small $w_{0,i}=-0.05+(1-i)/10$} &
\multicolumn{3}{|l|}{\small $f_i=0.4+1.6(i-1)/9$} \\
 & \multicolumn{3}{|l|}{\small $\Omega_{\Lambda,j}=-0.9+0.2(j-1)$} &  
\multicolumn{3}{|l|}{\small $w_{0,j}=-0.05-(j-1)/10$}
& \multicolumn{3}{|l|}{\small $w_{1,j}=-0.95(2j-11)/9$}
& \multicolumn{3}{|l|}{\small $M_j=2+(j-1)/3$} \\
\hline
Compared with & B & C & D & A & C & D & A & B & D & A & B & C \\
\hline
$P(\Delta\chi^2_{\rm min} < 11.8)$ & $19\,\%$ & $31\,\%$ 
& $14\,\%$ & $98\,\%$ & $93\,\%$ & $82\,\%$ & $95\,\%$ & $68\,\%$ & 
$40\,\%$ & $35\,\%$ & $23\,\%$ & $23\,\%$ \\ 
$P(\Delta\chi^2_{\rm min} < 6.17)$ & $15\,\%$ & $27\,\%$ 
& $9\,\%$ & $95\,\%$ & $93\,\%$ & $72\,\%$ & $79\,\%$ & $59\,\%$ & 
$26\,\%$ & $29\,\%$ & $20\,\%$ & $21\,\%$ \\
$P(\Delta\chi^2_{\rm min} < -11.8)$ & $0\,\%$ & $0\,\%$ & 
$0\,\%$ & $0\,\%$ & $0\,\%$ & $0\,\%$ & $0\,\%$ & $0\,\%$ & $< 1\,\%$ 
& $6\,\%$ & $7\,\%$ & $8\,\%$ \\
\hline
\end{tabular}
\begin{minipage}[t]{6.2in}
\caption{
Probabilities to obtain misleading results when the true model is model A, B, C 
and D at the $3\,\sigma$ level ($P(\Delta\chi^2_{\rm min} < 11.8)$) and at the 
$2\,\sigma$ level ($P(\Delta\chi^2_{\rm min} < 6.17)$). 
$P(\Delta\chi^2_{\rm min} < -11.8)$ is the probability to exclude the right 
model at the 
$3\,\sigma$ level. All these values are associated to the limits on the 
parameter spaces explored for generating the data.}
\label{tableresult}
\end{minipage}
\end{center}
\end{table*}

An important remark has to be made at this point\,: Figures  
\ref{figdotb}, \ref{figdotc} and \ref {figdotd} show that, whatever the true 
model is, a FLRW model with $\Omega_m = 0.3$ and $\Omega_\Lambda = 0.7$ ($w = -1$)
cannot be rejected on the basis of the SNAP observations. 

%
\subsection{Probability to select the correct model}

Table~\ref {tableresult} shows the probabilities to obtain misleading results when 
the true model is model A, B, C and D at the $3\,\sigma$ level ($P(\Delta\chi^2_{\rm min} 
< 11.8)$) and at the $2\,\sigma$ level ($P(\Delta\chi^2_{\rm min} < 6.17)$). It also 
presents the probability to exclude the right model at the $3\,\sigma$ level 
($P(\Delta\chi^2_{\rm min} < -11.8)$).
They give an idea of the degree of degeneracy between the models, with respect 
to the computation of the luminosity distances. 
For example, it is difficult to identify a true model B among wrong models\,: the 
probabilities to confuse model B with model A, C and D are respectively $95\,\%$, 
$93\,\%$ and $72\,\%$, at $2\,\sigma$ level. Model C cannot be distinguished from 
models A, B and D in $79\,\%$, $59\,\%$ and $26\,\%$ of the cases respectively 
($2\,\sigma$ level). Note that the use of model C as fiducial model assumed an 
exactly known $\Omega_m$. An uncertainty on $\Omega_m$ will increase 
the percentages given above. Finally, if the fiducial model is A or D, we can 
expect to rule out the proposed wrong models with an a priori probability higher 
than $65\,\%$ at the 3 $\sigma$ level. However this does not preclude a priori a degeneracy 
between those models any other possible model.

Samples simulated with model D can lead $\Delta\chi^2_{\rm min}$ 
values less than $- 11.8$ ($\sim 7\,\%$), which 
means that the correct model can be excluded at the $3\,\sigma$ level and 
confirms the sensitivity of model D to the statistical noise. Table \ref 
{tableresult} also shows that, whatever the true model is (B, C or D), the 
probability to reject model A at $3\,\sigma$ is low. 

The probabilities presented in Table~\ref {tableresult} clearly depend on the 
regions of the different parameter spaces used for the data simulations. 
However, in order to obtain meaningful results, we chose the parameter ranges 
surrounding the physical part of the contours presented in Figure \ref 
{figperlmutter}. 

\section{Conclusions}
\label{summary}
Making use of four quintessence models, of which three are mathematically 
simple and one is coming directly from particle physics, we have first shown 
that the various cosmological models may predict apparent magnitudes for 
objects at given redshifts which differ from each other by less than 0.04 
mag till $z = 2$. These magnitude differences are thus small compared to the 
intrinsic spread of the SNeIa maximum luminosity (0.15 mag). This indicates that 
an unambiguous discrimination between the cosmological models will be difficult 
to reach, with cosmological tests only based on luminosity distances. 

Second, we have fitted these models to the present supernovae data and found 
equally good fits with the four models. Finally, we have explored the 
discriminatory power of future SNeIa data to constrain the dark energy 
properties. We have simulated a large number of SNAP data with each particular 
model and have re-analysed them with the four cosmological models. 
We have then compared the positions and the values of the $\chi^2_{\rm min}$ 
obtained using wrong models with those obtained with the fiducial model. This 
led us to confirm the more skeptical conclusions already made in some previous 
works on the subject (Barger and Marfatia 2001, Astier 2001, Maor et~al. 2001, 
2002)\,: some degeneracies between the curvature, the matter density and the 
equation-of-state of dark energy will be difficult to break at the $3\,\sigma$
level and even at the $2\,\sigma$ and $1\,\sigma$ level. We also found that 
different estimates of a basic parameter like $\Omega_m$ can be obtained, 
depending on the model used 
for the data processing. Moreover, whatever the true model is, the presently 
admitted FLRW model with $\Omega_m = 0.3$ and $\Omega_\Lambda = 0.7$ will not 
be rejected on the basis of the future SNAP observations alone. 

In order to compare results expected from SNAP with the ones which could be 
obtained by other supernovae surveys, we have also generated data samples from 
models C and D for another redshift distribution\,: 2050 SNeIa between $z=0$ 
and $z=0.5$ and 50 between $z=2$ and $z=2.5$ (with $\sigma_i$ = 0.15 mag). 
This distribution reflects future observations to be made with the 4m Liquid 
Mirror Telescope (see ILMT URL) and with the Next Generation Space Telescope 
(see NGST URL). No significant difference has been found with the results 
presented here.

Many recent papers discussed the strong constraints that could be expected
from future SNeIa data on cosmological parameters. However, the most
important question is\,: how can we be sure that we consider the right
model~? Indeed how useful are strong constraints on cosmological parameters
if they do not describe the model chosen by Mother Nature~? The degeneracy 
problem presented in this paper is expected to affect every cosmological 
test using luminosity distance. Therefore it will be very difficult for 
SNAP alone to obtain any strong and safe constraints on the cosmological 
parameters. However, we expect that the combination of the CMB data with 
the future SNAP ones could help to break some of the luminosity distance 
degeneracies. This fact is already known in the framework of a single 
cosmological model (see e.g. Huterer \& Turner 2001, Frieman et al. 2002, 
Bean \& Melchiorri 2002). 

\section*{Acknowledgments}

The authors are grateful to E. Linder for ``stimulating'' comments on the very 
first version of this paper. It is also a pleasure to thank E. Gosset and A. Smette 
for constructive discussions and B. Gerke, A. Melchiorri and D. Wiltshire for 
drawing our attention to their recent works on a similar subject. This work was 
supported in part by Belgian Interuniversity Attraction Pole P5/36, by a grant from 
the ``Fonds National de la Recherche Scientifique'' (F.R.I.A.), and by PRODEX (Office
of Science, Technology, and Culture, Brussels).

\label{lastpage}
\end{document}